\documentclass[letterpaper]{article} 
\usepackage{aaai2026}  
\usepackage{times}  
\usepackage{helvet}  
\usepackage{courier}  
\usepackage[hyphens]{url}  
\usepackage{graphicx} 
\usepackage{natbib}  
\usepackage{caption} 
\usepackage{algorithm}
\usepackage{algorithmic}

\usepackage{subcaption}
\usepackage{tikz}
\usepackage{multirow}
\usepackage{tabularx}
\usetikzlibrary{quantikz2}
\usepackage[capitalise]{cleveref}
\usepackage{booktabs}
\usepackage{blochsphere}
\usepackage{tablefootnote}
\creflabelformat{equation}{#2#1#3}

\def\BibTeX{{\rm B\kern-.05em{\sc i\kern-.025em b}\kern-.08em
    T\kern-.1667em\lower.7ex\hbox{E}\kern-.125emX}}
\usepackage{newfloat}
\usepackage{listings}
\DeclareCaptionStyle{ruled}{labelfont=normalfont,labelsep=colon,strut=off} 
\floatstyle{ruled}
\newfloat{listing}{tb}{lst}{}
\floatname{listing}{Listing}
\title{Quantum Architecture Search for Solving Quantum Machine Learning Tasks}
\author {
    Michael Kölle, Simon Salfer, Tobias Rohe, Philipp Altmann, Claudia Linnhoff-Popien 
}
\affiliations {
    Institute of Informatics, LMU Munich, Munich, Germany\\
    michael.koelle@ifi.lmu.de
}

\begin{document}

\maketitle

\begin{abstract}
Quantum computing leverages quantum mechanics to address computational problems in ways that differ fundamentally from classical approaches. While current quantum hardware remains error-prone and limited in scale, Variational Quantum Circuits offer a noise-resilient framework suitable for today's devices. The performance of these circuits strongly depends on the underlying architecture of their parameterized quantum components. Identifying efficient, hardware-compatible quantum circuit architectures---known as Quantum Architecture Search (QAS)---is therefore essential. Manual QAS is complex and error-prone, motivating efforts to automate it. Among various automated strategies, Reinforcement Learning (RL) remains underexplored, particularly in Quantum Machine Learning contexts. This work introduces RL-QAS, a framework that applies RL to discover effective circuit architectures for classification tasks. We evaluate RL-QAS using the Iris and binary MNIST datasets. The agent autonomously discovers low-complexity circuit designs that achieve high test accuracy. Our results show that RL is a viable approach for automated architecture search in quantum machine learning. However, applying RL-QAS to more complex tasks will require further refinement of the search strategy and performance evaluation mechanisms.
\end{abstract}

\begin{links}
    \link{Code}{https://github.com/916750/qas4ml}
\end{links}

\section{Introduction}
\label{sec:introduction}

Quantum Computing (QC) is a computational paradigm that leverages the principles of quantum mechanics, offering fundamentally different mechanisms than classical computing. By exploiting quantum phenomena such as superposition, entanglement, and interference, QC is expected to deliver significant performance improvements—referred to as quantum advantage—in selected problem domains. These improvements may manifest as exponential speedups or reduced resource requirements. Among various QC models, circuit-based QC has gained prominence. In this model, quantum gates are composed into quantum circuits analogous to logical gates in classical systems.
Variational Quantum Circuits represent a particularly promising approach within circuit-based QC, especially suitable for current Noisy Intermediate-Scale Quantum (NISQ) devices. VQCs exhibit relative robustness to noise and hardware imperfections. At the core of a VQC lies a Parameterized Quantum Circuit (PQC), comprising tunable quantum gates. Similar to Artificial Neural Networks (ANNs), VQCs are trained by optimizing parameters to minimize a cost function. Their performance, however, is highly sensitive to the design of the PQC Architecture (PQCA), also known as the Ansatz.

Designing expressive, trainable, and hardware-compatible PQCAs is a critical challenge. The field of QAS has emerged to address this task. Manual QAS is resource-intensive, requiring interdisciplinary expertise and entailing high complexity and error risk. Consequently, automated QAS approaches have been explored using various search strategies. For instance, Evolutionary Algorithms have been applied to the Variational Quantum Eigensolver (VQE) problem~\cite{rattew2019domain,wang2022quantumnas,chivilikhin2020mog,huang2022robust}, while Differentiable QAS methods have been evaluated for Quantum Approximate Optimization Algorithm (QAOA) tasks~\cite{wu2023quantumdarts,zhang2022differentiable}. Monte Carlo Tree Search has also been studied in the context of Error Detection~\cite{wang2023automated}, VQE~\cite{meng2021quantum,wang2023automated}, and QAOA~\cite{wang2023automated,yao2022monte}.

Although these methods demonstrate potential, the QAS problem remains far from solved. The large search space of PQCAs and the computational cost of performance evaluation are persistent challenges that necessitate more efficient strategies.
RL is a promising but underexplored approach for QAS. Initial studies have applied RL-based QAS (RL-QAS) to the QAOA~\cite{mckiernan2019automated}, VQE~\cite{ostaszewski2021reinforcement,patel2024curriculum}, and Variational Quantum State Diagonalization~\cite{kundu2024enhancing} problems. These results indicate that RL-QAS can be effective. However, further investigation is required to assess its suitability across broader problem classes.

\begin{figure*}[htb]
  \centering
  \includegraphics[width=\linewidth]{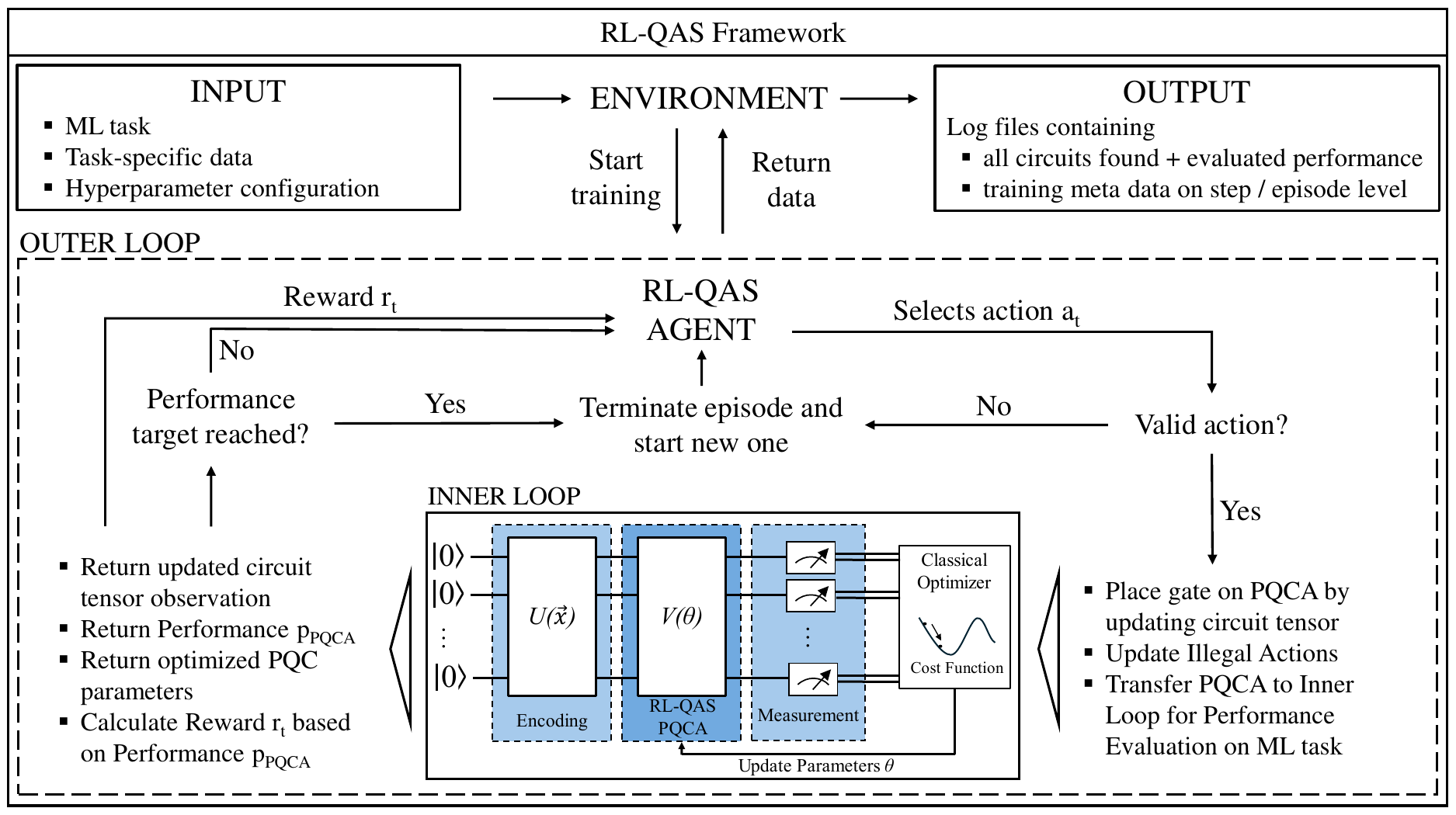}
  \caption[RL-QAS Framework Overview]{Overall architecture of the RL-QAS framework. The outer loop constructs PQCAs; the inner loop evaluates their performance using a VQC.}
  \label{fig:rlqas}
\end{figure*}

This paper investigates RL as a search strategy for QAS within the context of Quantum Machine Learning (QML). We propose a novel RL-QAS framework that decouples the search for PQCAs and their performance evaluation via a two-loop structure. In the outer loop, an RL agent constructs candidate PQCAs. In the inner loop, these circuits are trained and evaluated for a given task.
To assess this framework, we apply it to classification tasks using the Iris and binary MNIST datasets. Accuracy is used as the primary performance metric. We analyze the discovered PQCAs for structural patterns and performance characteristics. The results provide new insights into effective circuit architectures and validate RL as a viable strategy for QAS in QML.
This work contributes to both the QAS and QML domains by extending RL-QAS research to ML classification tasks, which remain largely unexplored.

The remainder of this paper is structured as follows.  
First, we detail prior work in quantum architecture search with a focus on classification in the section on Related Work. Next, we introduce our proposed approach in the section on RL-QAS Framework. The section on Experimental Setup outlines the benchmarks, implementation details, and evaluation criteria. We then present and analyze the empirical results in the section on Results.  
Finally, the sections on Discussion and Conclusion offer broader insights and summarize the main contributions of this work.

\section{Related Work}
\label{sec:related_work}

QAS is an emerging field aimed at automating the design of PQCAs. While early efforts focused on domains such as QAOA \cite{zhang2022differentiable,mckiernan2019automated} and VQE \cite{patel2024curriculum,ostaszewski2021reinforcement}, the application of QAS to machine learning (ML) classification tasks remains underexplored.
Several QAS methods for classification problems have been proposed using a variety of search strategies. Evolutionary Algorithms have shown promise in discovering efficient PQCAs with reduced depth and gate count. For instance, the Markovian Quantum Neuroevolution approach achieved high accuracy with significantly lower complexity for binary MNIST \cite{lu2021markovian}. Similarly, QuantumNAS integrated noise-aware and hardware-adaptive constraints into the search process and demonstrated superior performance across several QML benchmarks \cite{wang2022quantumnas}.
Predictor-based approaches like Neural Predictors \cite{zhang2021neural} and Graph Self-Supervised QAS \cite{he2023gsqas} accelerate the search by learning to estimate circuit performance, reducing reliance on costly simulations. Differentiable QAS strategies, such as QuantumDARTS \cite{wu2023quantumdarts}, adapt classical neural architecture search techniques to the quantum domain.
Despite these advances, RL has seen limited application in QAS for ML tasks. Prior work has applied RL to QAOA \cite{mckiernan2019automated}, VQE \cite{ostaszewski2021reinforcement}, and state discrimination problems \cite{kundu2024enhancing}, but not to classification tasks. This thesis addresses this gap by evaluating RL as a search strategy for discovering efficient PQCAs in the context of quantum classification, contributing to both QAS and QML research.

\section{RL-QAS Framework}
\label{sec:qas_framework}

This section introduces the RL-QAS framework developed in this work. It outlines the conceptual architecture of the system and its main components. Details regarding the specific experimental setup—including the classification task, encoding scheme, hyperparameters, and implementation—are provided in the experimental setup section. The framework is based on a Markov Decision Process, whose components have been adapted to the QAS problem and are discussed below.

\subsection{Observation Space}
\label{sec:observation_space}

Following \cite{patel2024curriculum}, each observation is encoded as a three-dimensional binary tensor representing the current PQCA. The VQC encoding and measurement stages are fixed and excluded from the observation, as they are not controlled by the RL-QAS agent. The tensor dimensions are given by $[Q \times (G + Q - 1) \times D]$, where $Q$ is the number of qubits, $G$ the size of the gate set, and $D$ the maximum circuit depth. The $(G + Q - 1)$ term accounts for the CNOT gate, which requires encoding both control and target qubits.

\begin{figure}[hpbt]
  \centering
  \includegraphics[width=\linewidth]{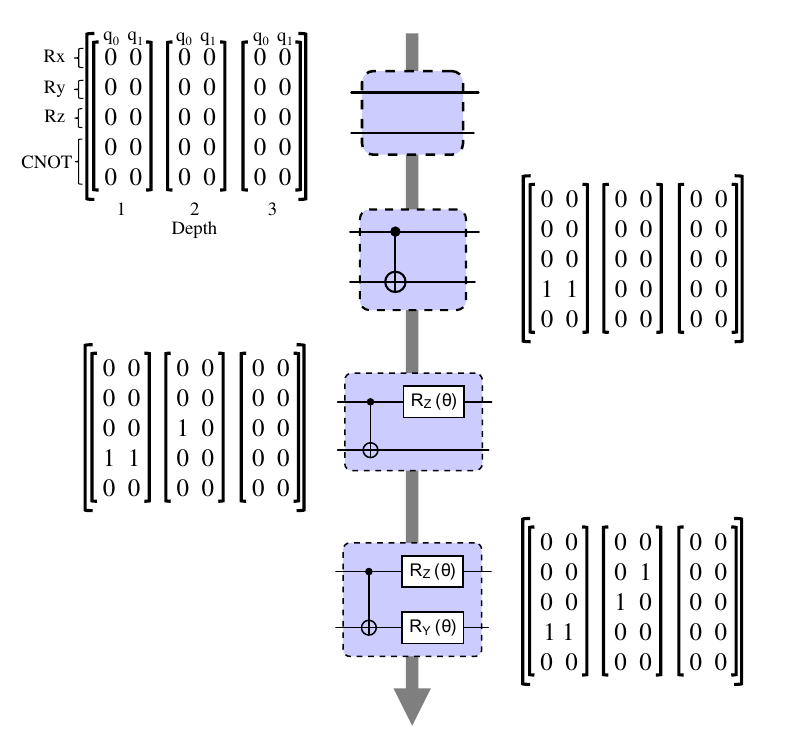}
  \caption[Example of tensor-based PQCA encoding]{Tensor-based PQCA encoding with a binary 3D tensor of size $[2 \times 5 \times 3]$ for a 2-qubit PQCA with gate set $\{R_x, R_y, R_z, \text{CNOT}\}$ and maximum depth 3. Adapted from \cite{kundu2024reinforcement}.}
  \label{fig:tensor_encoding_example}
\end{figure}

All tensor elements are initialized to zero at the start of an episode, corresponding to an empty PQCA. When the agent places a gate, the tensor is updated at the index defined by the triplet $(q, g, d)$, with $q$ denoting the qubit, $g$ the gate, and $d$ the depth level. This representation allows for all-to-all connectivity, supports arbitrary gate placements, and simplifies operations such as masking illegal actions. An example is illustrated in \cref{fig:tensor_encoding_example}.

\subsection{Action Space}
\label{sec:action_space}

The action space $\Gamma$ is defined as a multidiscrete set, following \cite{kolle2024reinforcement}. Each action $a \in A$ is a pair $(g_{\text{idx}}, q_{\text{idx}})$, where $g_{\text{idx}}$ selects a gate from the gate set $G = \{R_x, R_y, R_z, \text{CNOT}\}$, and $q_{\text{idx}}$ indexes the target qubit(s). Permutations of qubit indices are used to handle multi-qubit gates. For CNOT gates, the permutation includes all ordered qubit pairs. The number of such permutations $C$ is calculated as:
\begin{equation}
C = \frac{n!}{(n - n_{\text{max}})!}
\label{action_space}
\end{equation}
where $n$ is the number of qubits, and $n_{\text{max}}$ is the number of inputs for the most complex gate, here $n_{\text{max}} = 2$ for CNOT.
This leads to an action space of size $|G| \times |C|$, which grows quadratically with $n$. To reduce this size and improve training efficiency, we apply an illegal action mechanism based on \cite{kundu2024reinforcement}. Actions that violate defined constraints are pruned during training. Two primary constraints are: (i) placing the same gate consecutively on the same qubit, and (ii) exceeding the maximum depth on any qubit. Violations trigger early termination of the episode and incur a small penalty.



\subsection{Reward Shaping}
\label{sec:reward_shaping}

Reward shaping plays a central role in training the RL-QAS agent. The reward function comprises two equally weighted components: performance and complexity.
The performance component evaluates the PQCA’s effectiveness in solving the task—measured by accuracy in this study. The complexity component penalizes excessive circuit depth and gate count. Since the number of qubits is fixed, complexity $\text{C}_{\text{rem}}$ is computed from the number of gates and the depth only:
\begin{equation}
\label{complexity_component}
\begin{aligned}
\text{C}_{\text{rem}} &= \frac{\text{Depth}_{\text{remaining}} + \text{Gates}_{\text{remaining}}}{2}
\end{aligned}
\end{equation}

The total reward $r$ is computed as:
\begin{equation}
\label{total_reward}
\resizebox{\linewidth}{!}{$
r =
\begin{cases} 
0.1 \cdot \left( \frac{1}{2} P_{\text{delta}} + P_{\text{delta}}(C_{\text{rem}} + E_H) \right) & \text{if } a \notin I_A \wedge p < T_p \\
r_{LA} + 100 & \text{if } a \notin I_A \wedge p \geq T_p \\
-0.01 & \text{if } a \in I_A
\end{cases}
$}
\end{equation}

The performance delta $P_{\text{delta}}$ is defined as:
\begin{equation}
\label{performance_delta}
r =
\begin{cases} 
P_{\text{current}} & \text{if first action} \\
P_{\text{current}} - P_{\text{previous}} & \text{otherwise}
\end{cases}
\end{equation}

An extended horizon term $E_H = \text{Depth}_{\text{max}} \cdot 10$ ensures that later actions in an episode are weighted more heavily. 



\subsection{Training Loop}
\label{sec:training_loop}

An overview of the complete RL-QAS framework is shown in \cref{fig:rlqas}. At the beginning of training, the ML task and dataset are defined. Hyperparameters, including PPO configurations and maximum circuit depth, are then specified. A full list of tunable parameters is provided in \cref{tab:hyparams}.

Training is structured into an outer loop (architecture construction) and an inner loop (performance evaluation), following the approach in \cite{altmann2024challenges}. Each episode begins with an empty PQCA. The RL agent selects an action. If the action is illegal, the episode ends. If it is valid, the action updates the PQCA tensor, and the modified circuit is evaluated.

The inner loop trains the VQC using a classical optimizer and returns performance metrics. These are used to compute the reward, which is passed back to the agent. If the performance target is reached, the episode terminates. Otherwise, the agent continues adding gates until the depth limit is hit or another terminal condition is met.

\section{Experimental Setup} 
\label{sec:experimental_setup}

This section describes the experimental setup used to evaluate the RL-QAS framework introduced in the previous section. The experiments are limited to the machine learning task of classification. Exploring additional problem domains is left for future work.

\subsection{Datasets and Encoding}
\label{sec:datasets_preprocessing_encoding_method}

Two datasets were selected for classification: the Iris dataset and a binary subset of MNIST containing only the digits 0 and 1, as provided by Scikit-learn. Iris was used as a simple classification task with low dimensionality. To simplify the task further, all three binary class combinations of the original Iris dataset were considered before evaluating the full three-class version. Binary MNIST was used to assess the framework on a higher-dimensional, more complex task.

All data were preprocessed for compatibility with VQCs. Features were normalized using the L2 norm to enable amplitude encoding. Labels were one-hot encoded to facilitate class assignment during measurement post-processing. Each dataset was split into 70\% training and 30\% testing subsets. For MNIST, dimensionality was reduced from 64 to 32 using Principal Component Analysis (PCA), retaining 97.6\% of the variance.



Amplitude encoding was used for all experiments due to its compactness, which helps minimize the number of qubits required. For Iris (4 features), 2 qubits were sufficient. For MNIST (32 features post-PCA), 5 qubits were needed. PCA reduced the required qubits from 6 to 5, shrinking the RL agent's action space from $4 \times 30 = 120$ actions to $4 \times 20 = 80$ actions.

\subsection{Measurement and Post-processing}
\label{sec:measurement_postprocessing_and_performance_metric}

During measurement, computational basis state probabilities are extracted. The final prediction is made using the argmax strategy—selecting the class associated with the most probable quantum state. Depending on the number of classes and qubits measured, basis states are evenly mapped to class labels.
Accuracy was chosen as the performance metric, as both datasets exhibit balanced class distributions. To promote generalization, the reward is calculated based solely on test accuracy. This encourages the RL-QAS agent to construct PQCAs that generalize well rather than overfitting the training data.

\subsection{Hyperparameter Tuning}
\label{sec:hyperparameter_tuning}

Hyperparameters of the PPO algorithm were prioritized for optimization due to their strong influence on training success. Automated tuning was not feasible due to the high computational cost of training. Instead, a manual grid search approach was adopted, varying key parameters such as the learning rate \([0.001,\ 0.003,\ 0.005]\) and the entropy coefficient \([0.01,\ 0.015,\ 0.02,\ 0.025,\ 0.03]\). Additional PPO hyperparameters included the number of steps (\texttt{n\_steps}) \([128,\ 512,\ 1024]\), and the batch size \([64,\ 128]\).
In the context of circuit parameter optimization, several parameter initialization ranges were explored: \([-0.5,\ 0.5]\), \([-1.0,\ 1.0]\), \([-2.0,\ 2.0]\), and \([-\pi,\ \pi]\). Maximum circuit depth was initially set to \(4\) for all tasks. For Iris, depths of \(5\) and \(6\) were also tested. For MNIST, depths of \(4\) to \(7\) were evaluated. The number of training steps was adjusted based on task complexity. In the inner loop, PQCAs were optimized using the Adam optimizer with a learning rate of \(0.01\). Each PQCA was evaluated across three independent runs using different random seeds. Parameter initialization was drawn uniformly from \([-1.0,\ 1.0]\). The cost function was cross-entropy loss. An overview of hyperparameter configurations is provided in \cref{tab:hyparams}.

\begin{table}[htb]
\centering
\footnotesize
\begin{tabularx}{\linewidth}{Xccc}
\toprule
\textbf{Hyperparameter} & \textbf{Iris 2} & \textbf{Iris} & \textbf{MNIST 2} \\
\midrule
Number of runs & 3 & 3 & 3 \\
Max. Circuit Depth & 4 & 4, 5, 6 & 4, 5, 6, 7 \\
Training Steps per Run & 100,000 & 200,000 & 400,000 \\
Learning Rate (PPO) & 0.003 & 0.003 & 0.003 \\
n steps & 128 & 512 & 1024 \\
Batch size & 128 & 128 & 128 \\
N epochs & 10 & 10 & 10 \\
Gamma & 0.99 & 0.99 & 0.99 \\
Gae lambda & 0.95 & 0.95 & 0.95 \\
Clip range & 0.2 & 0.2 & 0.2 \\
Ent coeff & 0.03 & 0.03 & 0.03 \\
VF coeff & 0.5 & 0.5 & 0.5 \\
Max grad norm & 0.5 & 0.5 & 0.5 \\
Net arch & [64, 64] & [64, 64] & [64, 64] \\
Optimizer & Adam & Adam & Adam \\
Learning Rate (Opt.) & 0.01 & 0.01 & 0.01 \\
Param. Init. Range & [-1.0, 1.0] & [-1.0, 1.0] & [-1.0, 1.0] \\
Param. Init. Prob. Distribution & Uniform & Uniform & Uniform \\
Batch Size (Opt.) & 16 & 20 & 20 \\
Opt. Runs within Inner Loop & 3 & 3 & 3 \\
Opt. Epochs & 1,000 & 1,000 & 1,000 \\
\bottomrule
\end{tabularx}
\caption{Hyperparameter configuration for Iris 2, Iris, and MNIST 2 experiments}
\label{tab:hyparams}
\end{table}

\begin{figure*}[htbp]
  \centering
  \subfloat[Test accuracy]{\includegraphics[width=0.24\linewidth]{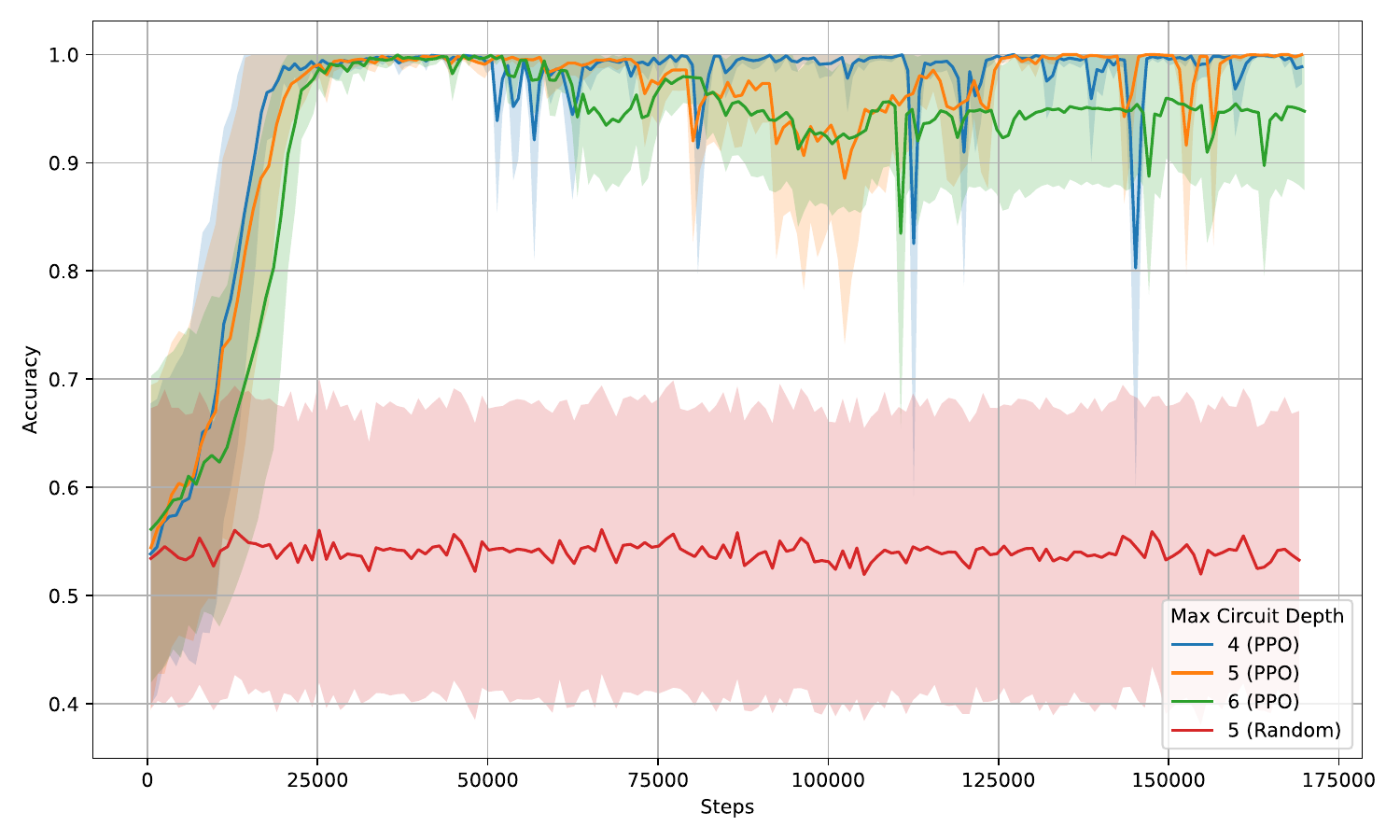}}\hfill
  \subfloat[Episode reward]{\includegraphics[width=0.24\linewidth]{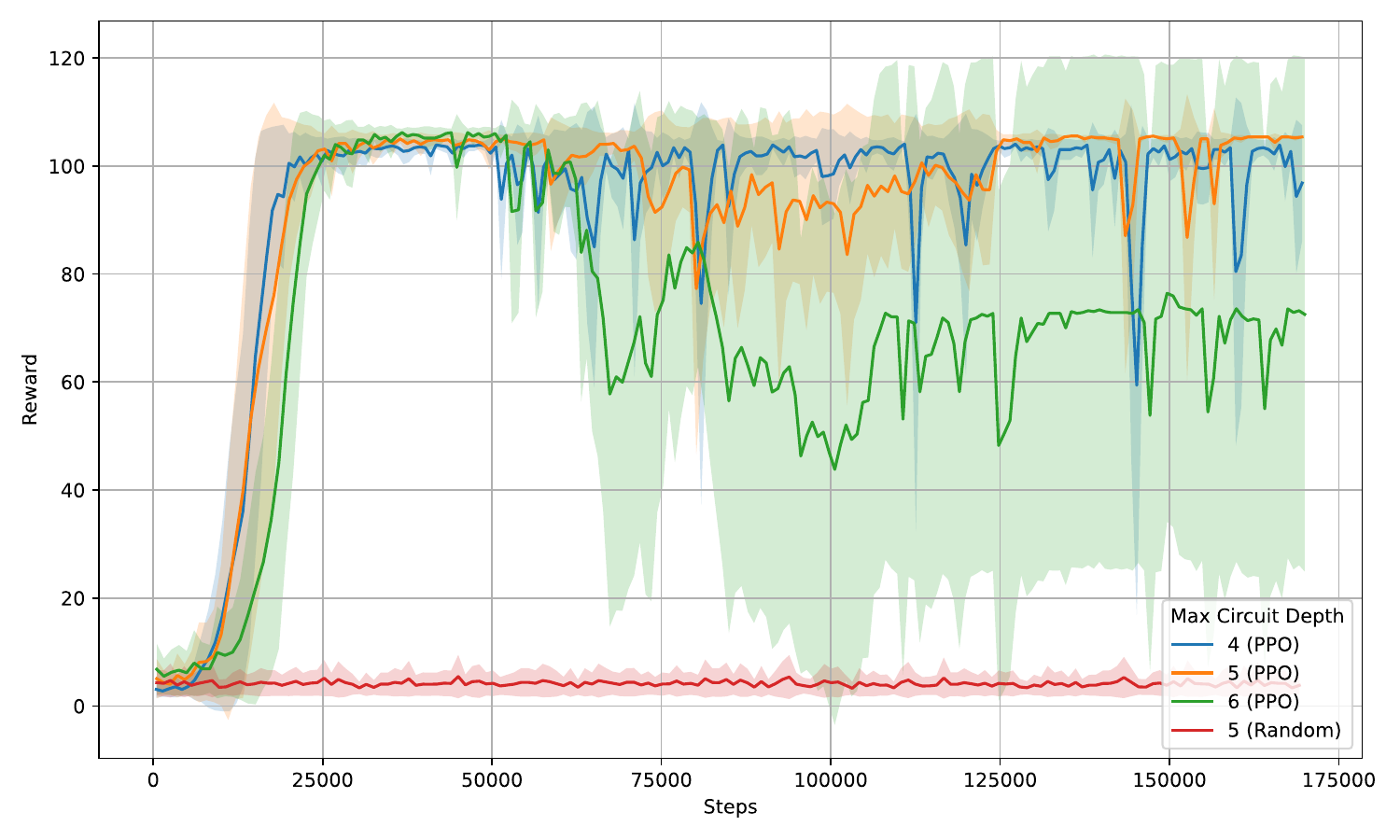}}\hfill
  \subfloat[Number of gates]{\includegraphics[width=0.24\linewidth]{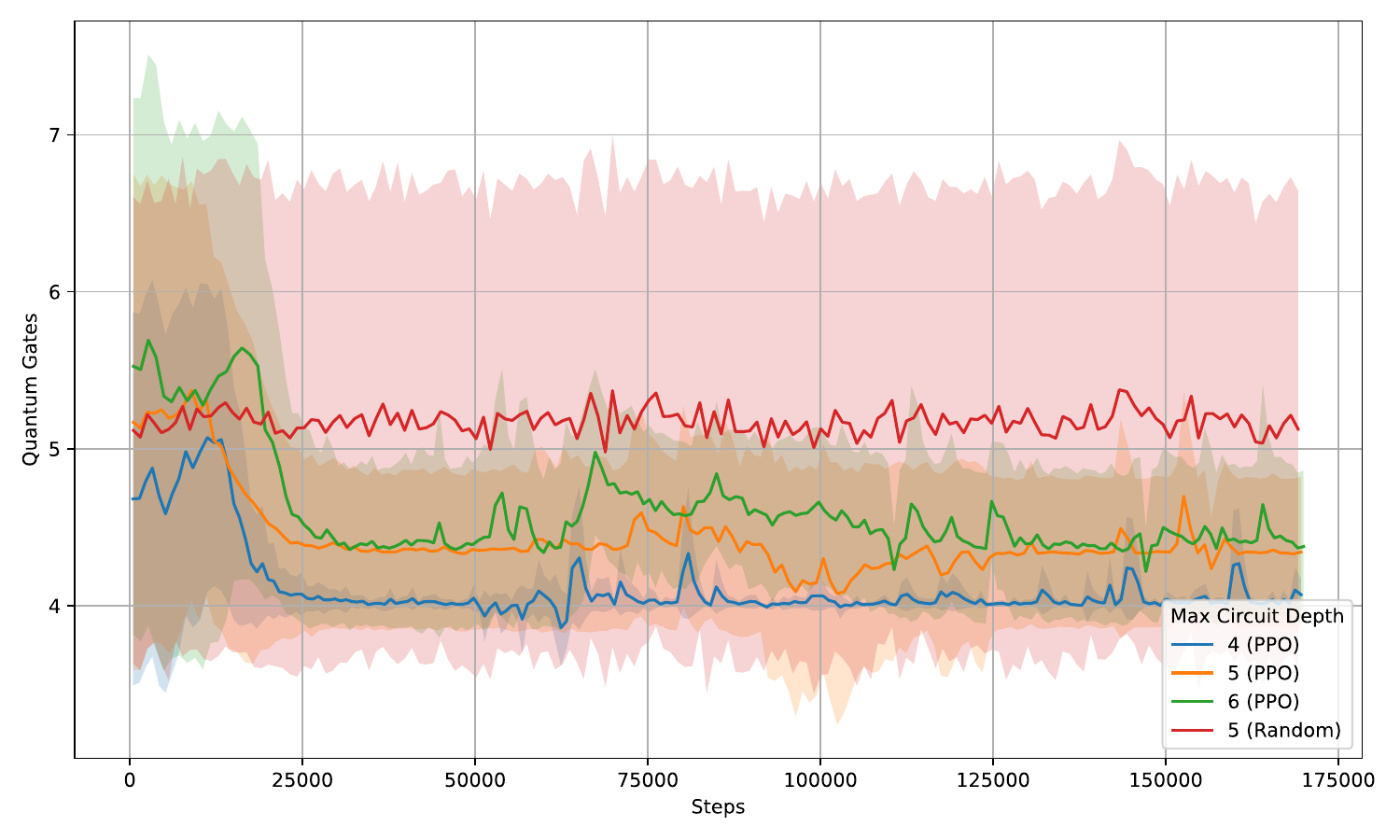}}\hfill
  \subfloat[Circuit depth]{\includegraphics[width=0.24\linewidth]{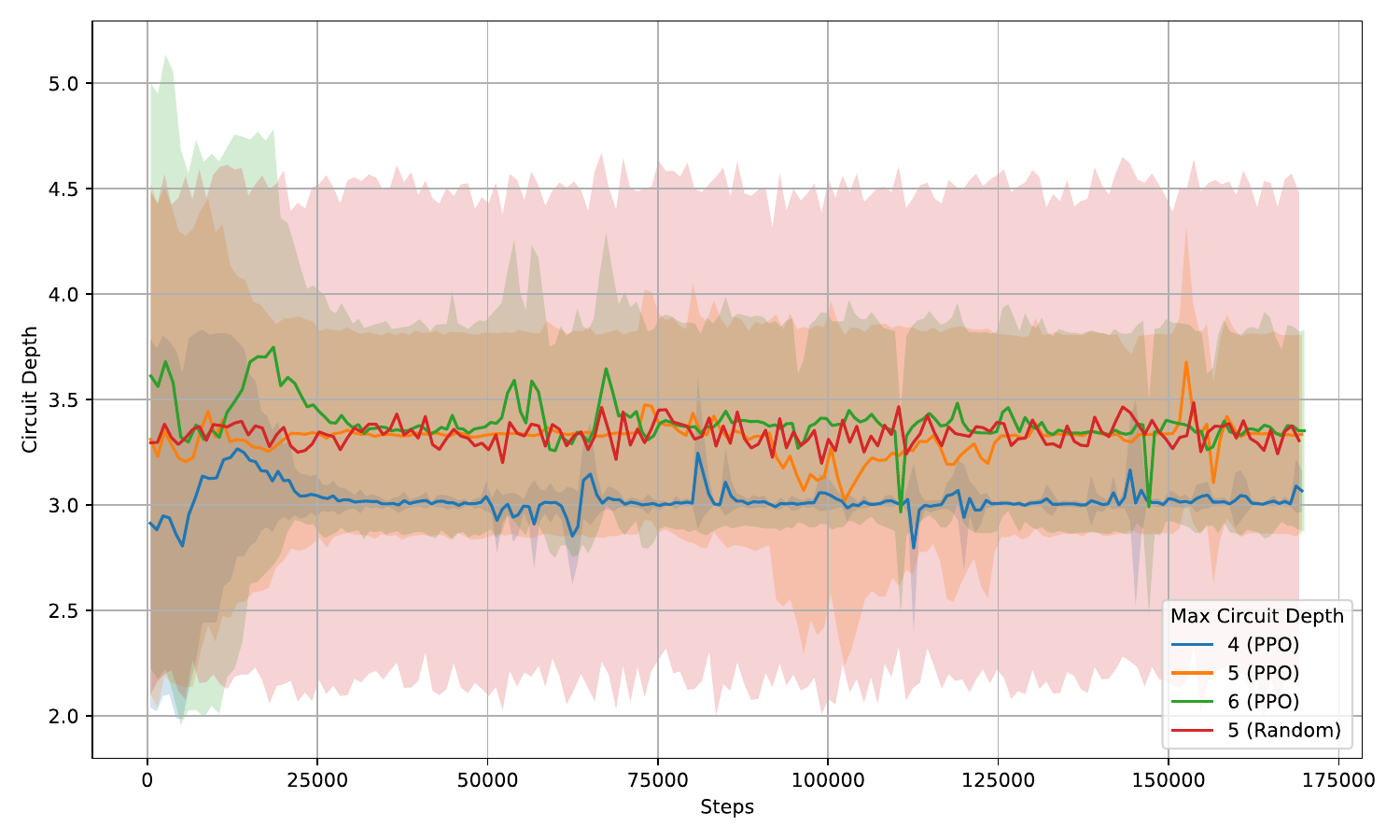}}
  \caption{Training performance of the RL-QAS agent for Iris using test accuracy, episode reward, number of gates and circuit depth.}
  \label{fig:accuracy_reward_x_steps_iris}
\end{figure*}

\begin{table*}[htb]
  \centering
  \includegraphics[width=\linewidth]{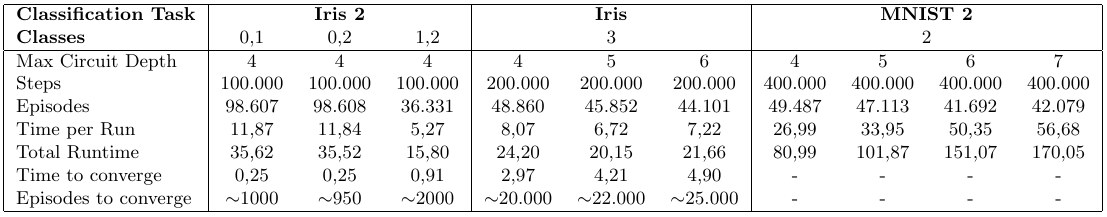}
  \caption[RL-QAS training metadata]{Training metadata for Iris 2, Iris, and MNIST 2. Step, episode, and time values are averages over 3 runs. Times are reported in hours. The MNIST 2 agent did not converge.}
  \label{tab:table_agent_runs_meta_data}
\end{table*}

\subsection{Benchmarking}
\label{sec:benchmarking}

To assess the effectiveness of RL-QAS, several baselines were used. First, a random agent served as a reference for the PPO-trained RL-QAS agent. This baseline selects actions uniformly at random and uses identical hyperparameters except those related to PPO.

\begin{figure}[htb]
  \centering
  \begin{minipage}[b]{0.49\linewidth}
    \centering
    \includegraphics[width=\linewidth]{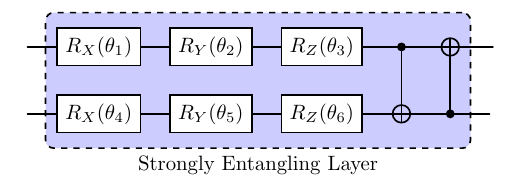}
    \subcaption{Iris}
  \end{minipage}
  \begin{minipage}[b]{0.49\linewidth}
    \centering
    \includegraphics[width=\linewidth]{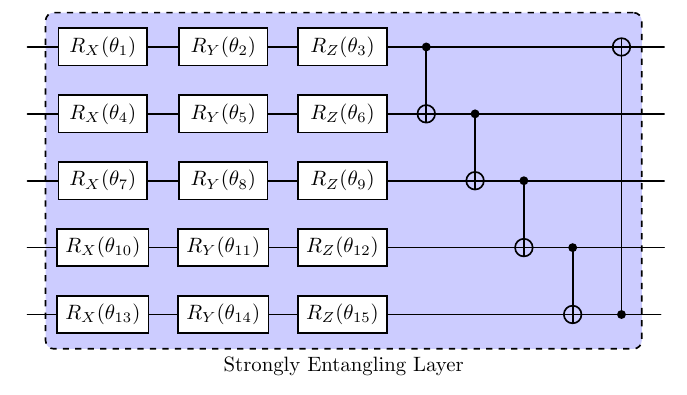}
    \subcaption{MNIST 2}
  \end{minipage}
  \caption[SEL PQCA architecture for benchmarking]{Architecture of the SEL PQCA used for benchmarking (a) Iris and (b) MNIST 2. Each subplot shows one SEL layer. Multiple layers can be stacked.}
  \label{fig:benchmark_sel_vqcs}
\end{figure}

Second, a Strongly Entangling Layer (SEL) VQC was used to benchmark final performance. The best-performing PQCA discovered by RL-QAS was compared with the SEL design in terms of accuracy, training dynamics, and complexity. The SEL architecture is illustrated in \cref{fig:benchmark_sel_vqcs} for both Iris and MNIST.

\subsection{Implementation Details}
\label{sec:technical_implementation_details}

The RL-QAS framework was implemented in Python using OpenAI Gym. PennyLane was used for quantum circuit modeling and simulation, with performance evaluation conducted on noise-free simulators. The PPO algorithm was implemented using the JAX-optimized Stable Baselines3 framework.
Scikit-learn was used for dataset loading and preprocessing. The inner loop—responsible for VQC simulation and parameter optimization—was accelerated using JAX Just-in-Time compilation.
To reduce redundant computations, a caching mechanism was developed. For each unique PQCA, a hash value (based on its tensor encoding) was generated and used as a key in a persistent hash map. If a previously evaluated PQCA reappeared, its stored performance was reused. This cache supports concurrent read/write access from multiple agents and persists across training sessions.

\section{Results} 
\label{sec:results}

This section presents the results obtained using the experimental setup described in the previous section. First, we analyze the RL-QAS agent's training progress using four key metrics—accuracy, reward, gate count, and circuit depth. Then, we evaluate the PQCAs discovered by the agent at a macro level, including descriptive statistics and recurring architecture patterns. Finally, we offer a micro-level analysis of the best PQCAs, including optimization behavior and cost landscape visualization. The focus is primarily on the Iris classification task.

\subsection{Performance of the RL-QAS Agent}
\label{sec:performance_of_the_rl_qas_agent}

\Cref{tab:table_agent_runs_meta_data} summarizes the training metadata for each classification problem. As expected, training duration, number of steps, and episodes required for convergence increase with task complexity. For MNIST 2, the agent fails to converge, indicating that further hyperparameter tuning is necessary.

For Iris 2 (classes 0 and 1), convergence is reached within 15 minutes and ~1,000 episodes. For the full Iris dataset, convergence occurs after ~3 hours and 20,000 episodes. These trends confirm that more complex tasks require longer training. Notably, caching was not used in early runs on Iris 2 (classes 0 and 1, and 0 and 2), resulting in runtimes nearly twice as long. This underscores the efficiency benefits of caching.
In MNIST 2, longer training and deeper circuits significantly increase runtime—from 81 hours (depth 4) to 170 hours (depth 7). Episode count is inversely related to circuit depth, as longer circuits allow more actions before episode termination. To assess learning progress, all four metrics—accuracy, reward, gate count, and depth—are evaluated per training step.

Plots for all Iris 2 and MNIST 2 runs are included in the supplementary material. For Iris 2 (classes 0 and 1 or 0 and 2), the agent quickly discovers a PQCA that achieves 100\% accuracy using a single gate, due to the linear separability of the classes. For Iris 2 (classes 1 and 2), the agent converges to 100\% accuracy in ~10,000 steps and later stabilizes at 96\% with only two gates. The reward function and illegal action mechanism appear effective—agents sometimes terminate early to avoid adding unnecessary gates.
In the Iris problem, the agent consistently reaches 100\% accuracy across all max depth configurations by 25,000 steps. However, training becomes slightly unstable beyond 50,000 steps, particularly at greater depth limits. The agent outperforms the random baseline in all metrics. The number of gates and circuit depth decrease as the agent refines its architecture.
Training on MNIST 2 was more unstable due to higher input dimensionality and possibly suboptimal hyperparameters. Still, performance improves with increased circuit depth, suggesting that deeper circuits enhance expressibility. The agent uses nearly all allowed depth, but not all gate slots, indicating a preference for deeper yet compact circuits. Additional pruning and efficiency mechanisms are likely required for this task.

\subsection{Macro-Analysis of Circuit Architectures}
\label{sec:macro_analysis_of_circuit_architectures}

This section analyzes architectural trends in the PQCAs discovered for Iris. Over 9,000 unique PQCAs were found—still a small fraction of the ~36 million theoretical designs. This demonstrates the efficiency of RL-based architecture search. Most PQCAs scored above 60\% accuracy, with the count increasing for higher accuracies and deeper circuits. This suggests the agent effectively leverages higher complexity to discover better architectures. Even circuits with only four gates and depth three achieve perfect accuracy. This validates the reward shaping approach, which favors efficient yet performant designs.

\begin{figure}[hpbt]
  \centering
  \subfloat[Unique PQCAs by test accuracy and circuit depth for the Iris classification problem.\label{fig:acc_unique_circuits_per_max_depth_bar_iris}]{
    \includegraphics[width=0.47\linewidth]{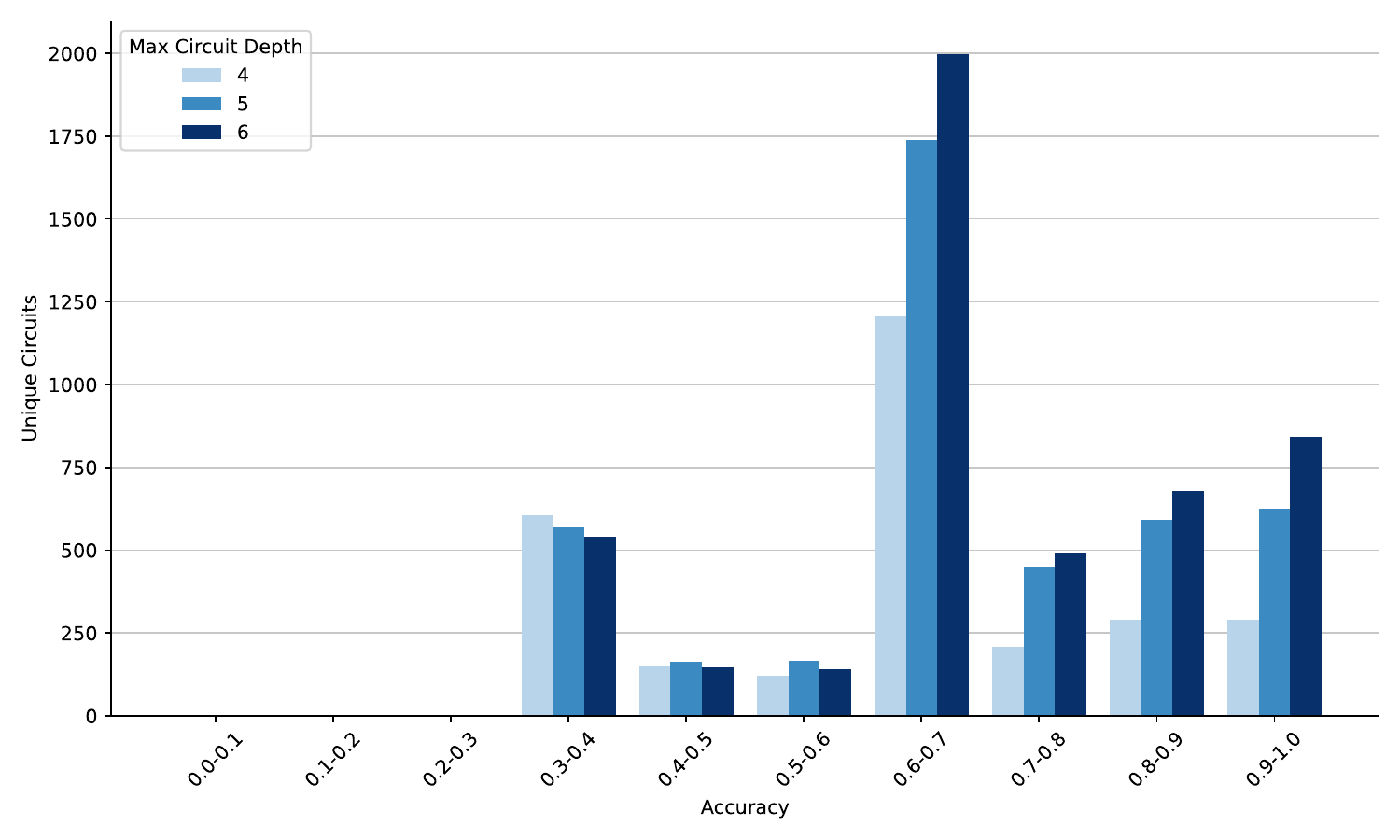}
  }
  \hfill
  \subfloat[Gate usage frequency across qubits in PQCAs with $\geq90\%$ test accuracy.\label{fig:gate_frequencies_plot_iris_cn1_pt_0.9}]{
    \includegraphics[width=0.47\linewidth]{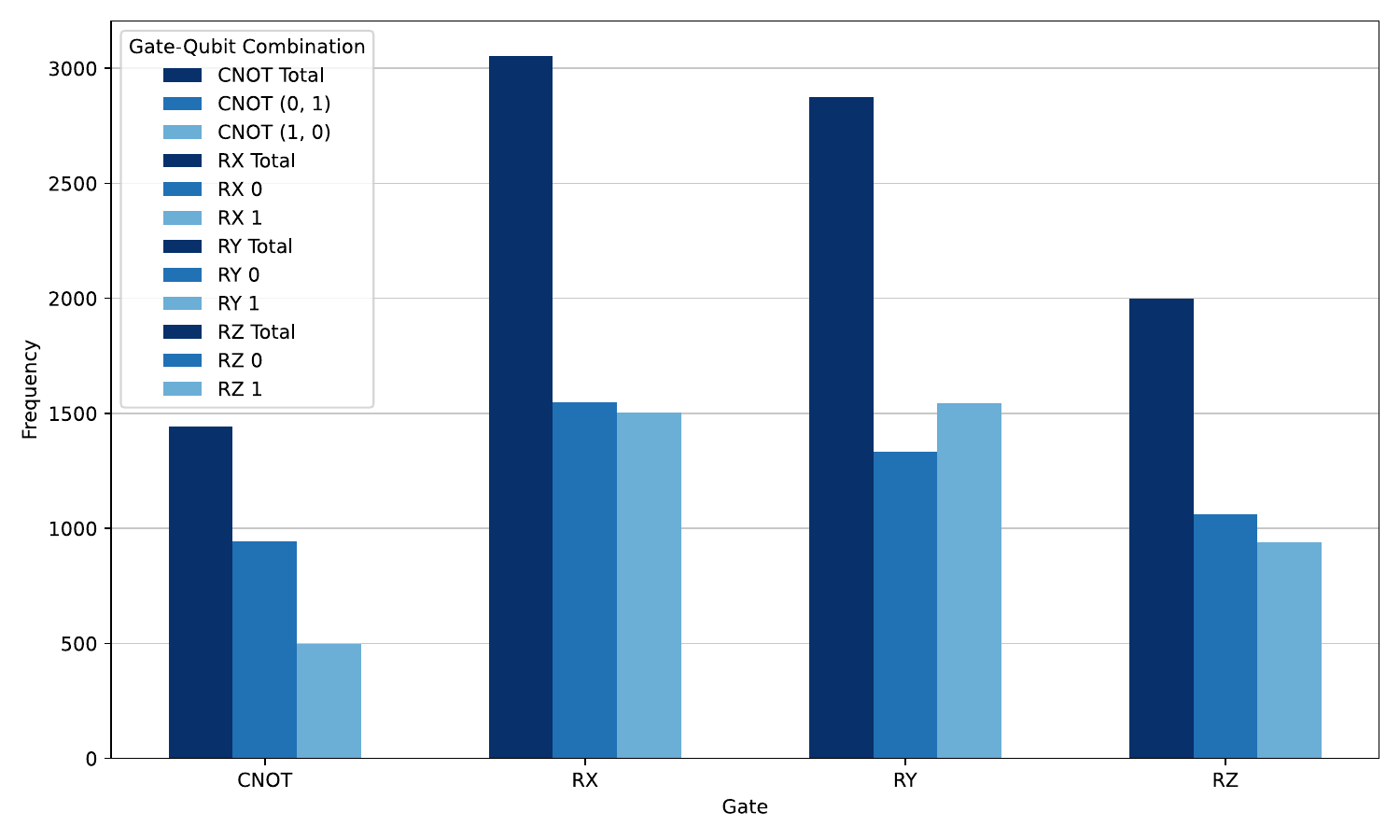}
  }
  \caption{Comparison of PQCAs in terms of test accuracy, circuit depth, and gate usage for the Iris classification task.}
  \label{fig:iris_pqca_comparison}
\end{figure}



\subsection{Recurring Architecture Patterns}
\label{sec:recurring_architecture_patterns}

To identify architectural trends, all PQCAs achieving at least 90\% test accuracy were analyzed. Rx and Ry gates dominate across sequences and qubits. CNOT gates are mostly used early in the circuits and often operate with qubit 0 as control and qubit 1 as target. Patterns such as \texttt{[CNOT, Ry, Rx, Ry]} appear frequently and exhibit symmetry. Depth-wise, CNOTs are used early, followed by rotation gates. Most PQCAs use 3–4 layers, with Rx and Ry dominating at greater depths.




\begin{figure}[hbt]
  \centering
  \subfloat[Heat map of transition probabilities for gate pairs in high-performing PQCAs.\label{fig:gate_sequences_heatmap_iris_cn1_pt_0.9}]{
    \includegraphics[width=0.47\linewidth]{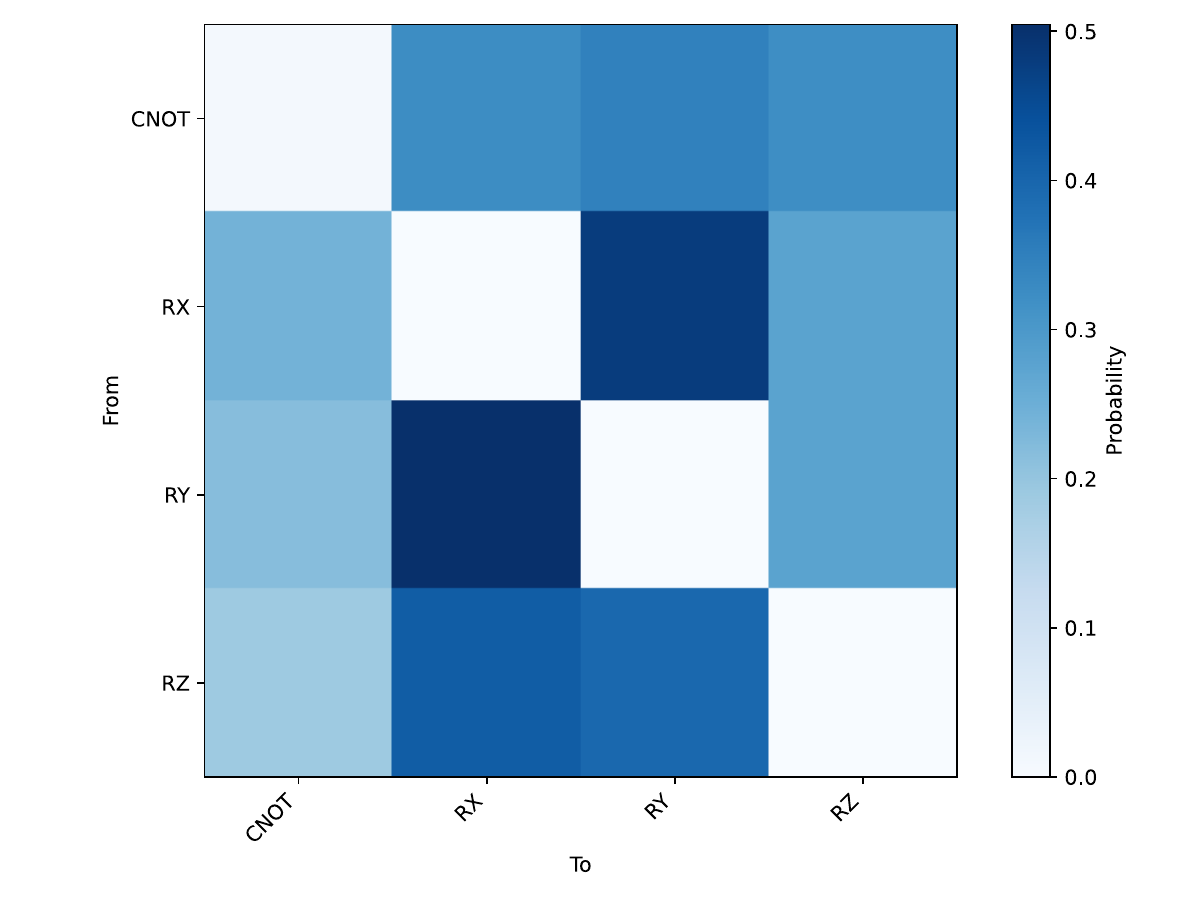}
  }\hfill
  \subfloat[Gate type distribution by circuit depth in high-performing PQCAs.\label{fig:gate_distribution_depth_heatmap_iris_cn1_performance_threshold_0.9}]{
    \includegraphics[width=0.47\linewidth]{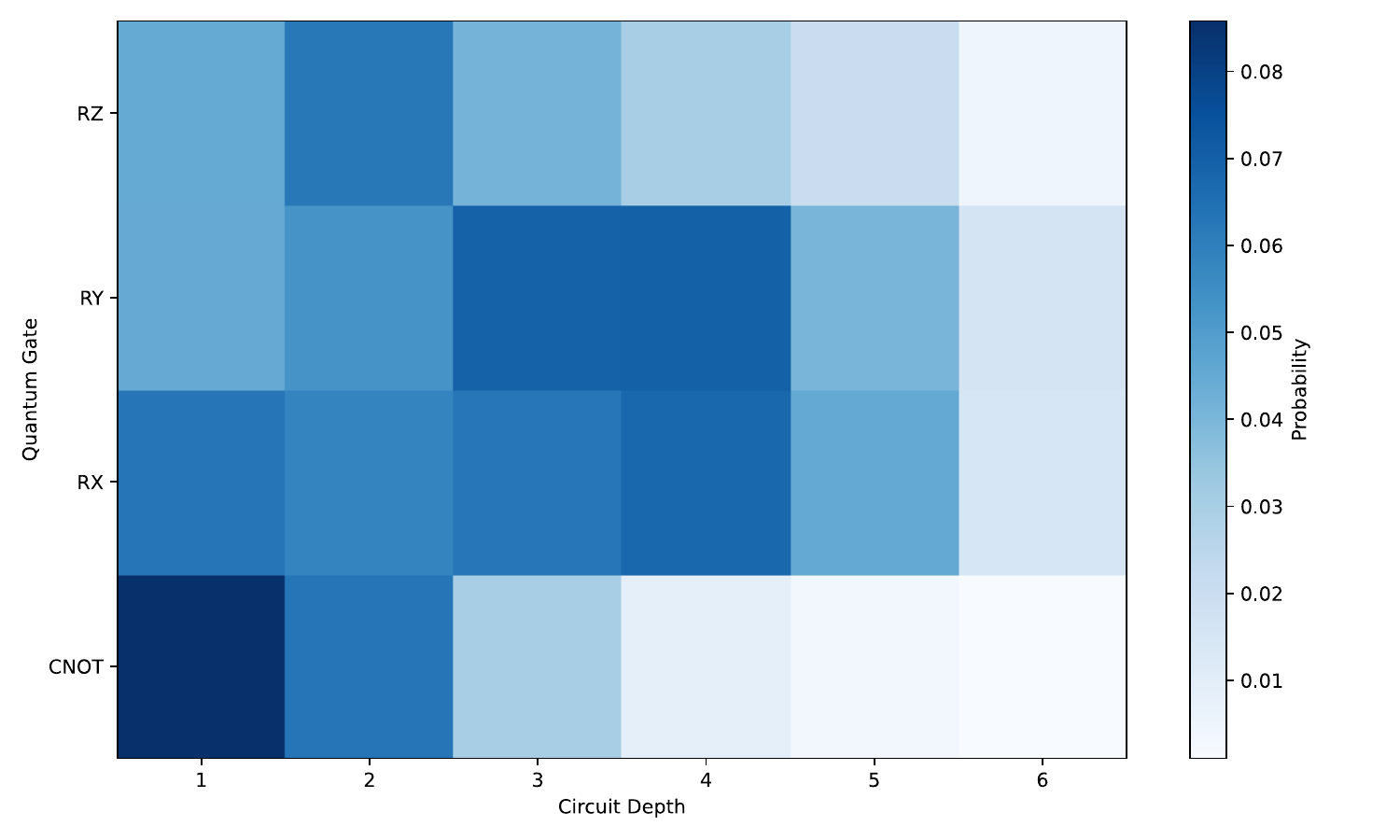}
  }
  \caption{Analysis of gate pair transitions and gate usage by depth in high-performing PQCAs for the Iris classification task.}
  \label{fig:gate_transition_depth_analysis}
\end{figure}


\subsection{Micro-Analysis of Circuit Architectures}
\label{sec:micro_analysis_of_circuit_architectures}

\begin{figure*}[htb]
  \centering
  \subfloat[Binary Iris\label{fig:best_pqcas_iris_2}]{
    \includegraphics[width=0.48\linewidth]{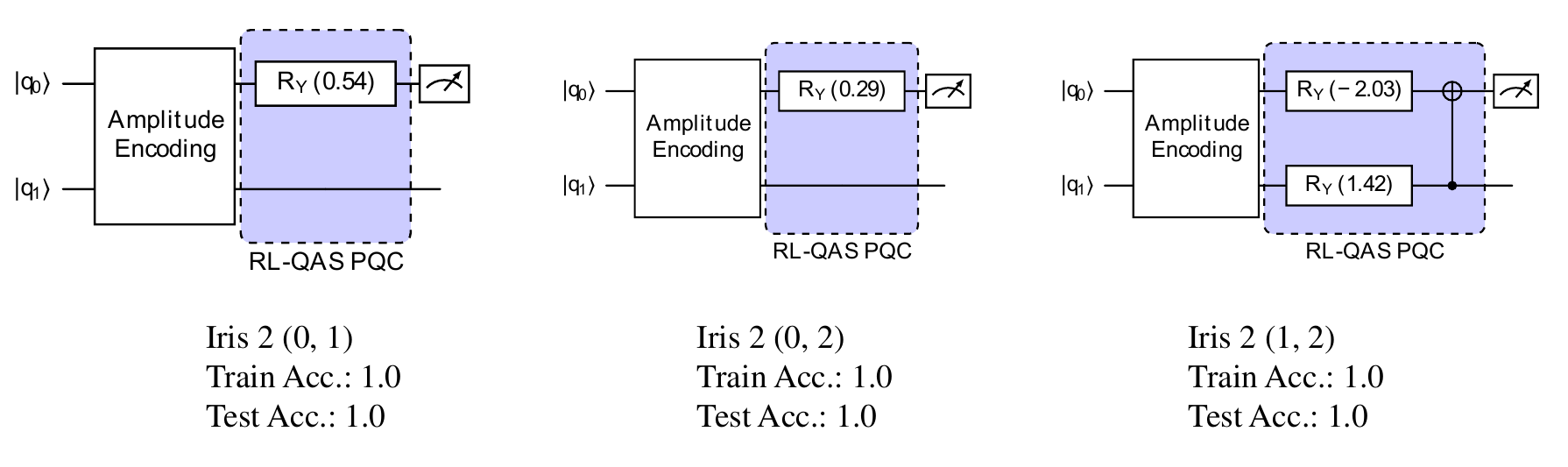}
  }%
  \hfill
  \subfloat[Iris\label{fig:best_pqcas_iris}]{
    \includegraphics[width=0.24\linewidth]{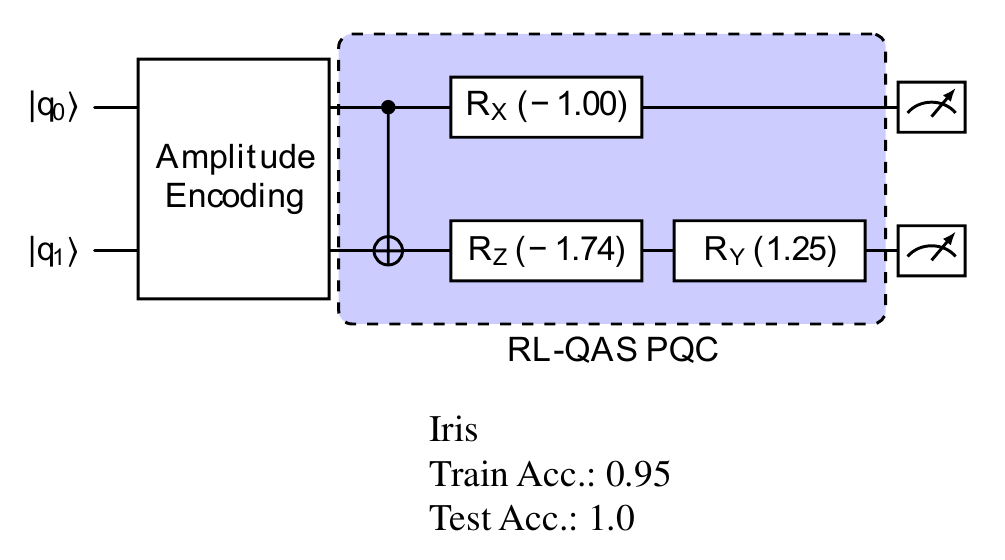}
  }%
  \hfill
  \subfloat[MNIST 2\label{fig:best_pqcas_mnist_2}]{
    \includegraphics[width=0.24\linewidth]{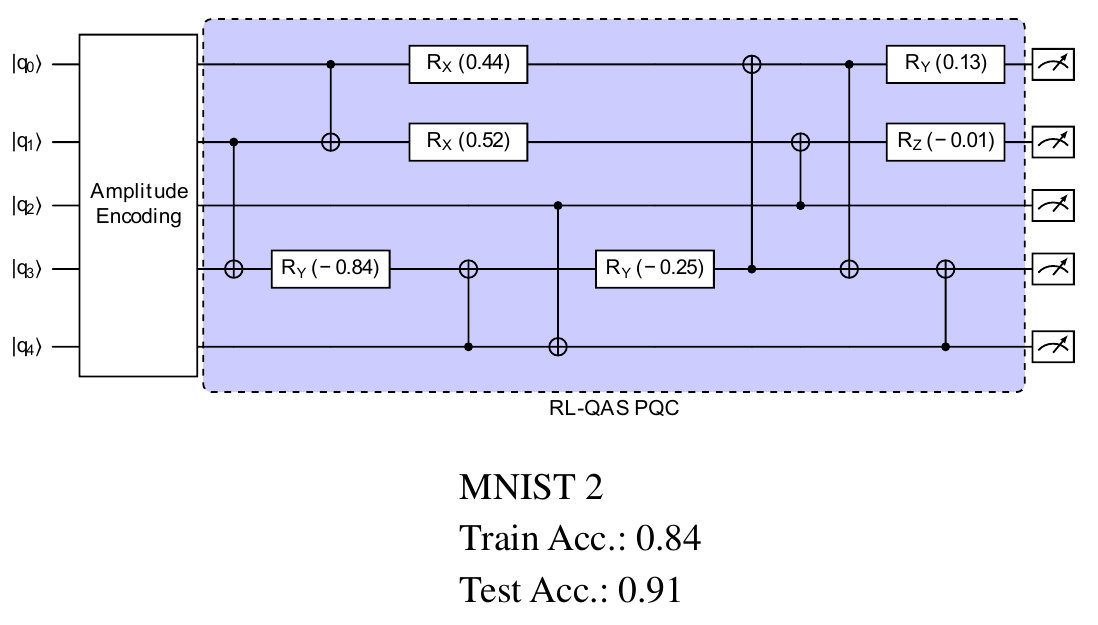}
  }
  \caption{Best PQCAs for binary Iris, Iris, and MNIST 2 classification.}
  \label{fig:best_pqcas_all}
\end{figure*}

\Cref{fig:best_pqcas_iris_2,fig:best_pqcas_iris} show the best PQCAs for the binary and full Iris problems. For linearly separable classes, a single Ry gate suffices. For non-separable classes, a CNOT and multiple rotations are used. The RL-QAS VQCs outperform SEL baselines with fewer gates. For MNIST 2, the best PQCA is more entangled and contains several CNOTs, shown in \cref{fig:best_pqcas_mnist_2}.

\begin{table}[htbp]
\centering
\small
\begin{tabularx}{\linewidth}{llXXXXXXX}
\toprule
\textbf{Task} & & \textbf{G} & \textbf{P} & \textbf{C} & \textbf{D} & \textbf{TrA} & \textbf{TeA} \\
\midrule
\textbf{Iris 2 (0,1)} & RL-QAS VQC & 1 & 1 & 0 & 1 & 1.0 & 1.0 \\
 & SEL VQC (1) & 8 & 6 & 2 & 4 & 0.98 & 1.0 \\
\textbf{Iris 2 (0,2)} & RL-QAS VQC & 1 & 1 & 0 & 1 & 1.0 & 1.0 \\
 & SEL VQC (1) & 8 & 6 & 2 & 4 & 0.98 & 1.0 \\
\textbf{Iris 2 (1,2)} & RL-QAS VQC & 3 & 2 & 1 & 2 & 1.0 & 1.0 \\
 & SEL VQC (1) & 8 & 6 & 2 & 4 & 0.98 & 1.0 \\
\textbf{Iris} & RL-QAS VQC & 4 & 3 & 1 & 3 & 0.95 & 1.0 \\
 & SEL VQC (1) & 8 & 6 & 2 & 4 & 0.66 & 0.66 \\
 & SEL VQC (2) & 16 & 12 & 4 & 8 & 0.84 & 0.82 \\
 & SEL VQC (3) & 24 & 18 & 6 & 12 & 0.73 & 0.77 \\
\textbf{MNIST 2} & RL-QAS VQC & 14 & 6 & 8 & 7 & 0.84 & 0.91 \\
 & SEL VQC (1) & 20 & 15 & 5 & 8 & 0.77 & 0.77 \\
 & SEL VQC (2) & 40 & 30 & 10 & 16 & 0.86 & 0.93 \\
\bottomrule
\end{tabularx}
\label{tab:tab_rlqas_sel_complexity_performance_comparison}
\caption{Comparison of RL-QAS and SEL VQCs across tasks with respect to complexity and performance metrics gates (G), parameters (P), CNOTs (C), circuit depth (D), training accuracy (TrA) and test accuracy (TeA).}
\end{table}

\begin{figure}[hpbt]
  \centering
  \includegraphics[width=0.8\linewidth]{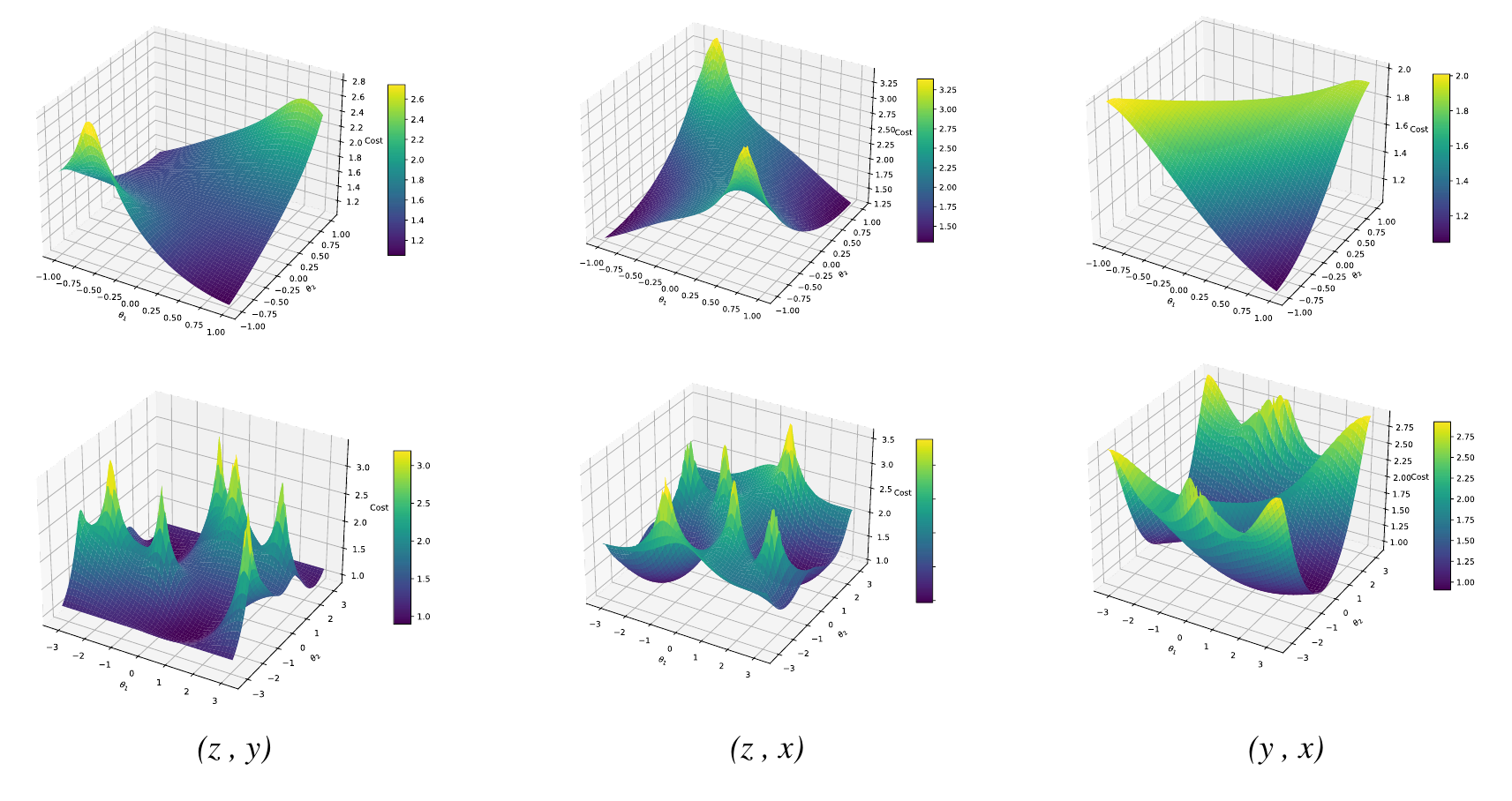}
  \caption[Cost landscape visualization for the Iris PQCA]{Cost landscape of the best PQCA for Iris across three parameter pairs in two parameter intervals.}
  \label{fig:cost_landscape_iris}
\end{figure}

\Cref{fig:cost_landscape_iris} visualizes the cost landscape for the Iris PQCA. The upper plots show a smooth surface with a global minimum in $[-1, 1]$, while the lower plots over $[-\pi, \pi]$ reveal a periodic, multimodal landscape. Parameter entanglement, especially due to the CNOT gate, necessitates coordinated optimization.

\section{Discussion}
\label{sec:discussion}

This section critically evaluates the results presented in the previous section, with reference to the primary objective of this work—assessing RL as a viable strategy for QAS. In addition to assessing the contributions, key limitations and assumptions are discussed, and directions for future research are proposed.

\subsection{Reinforcement Learning as a QAS Strategy}
\label{sec:rl_as_qas_strategy}

This work investigated the use of RL as a search strategy for identifying efficient PQCAs in the context of QAS. The RL-QAS framework was evaluated on two classification problems—\textit{Iris} and a binary variant of \textit{MNIST}—and the resulting PQCAs were further analyzed to gain insights into their structure and common patterns. These insights contribute to both QAS and QML research.

Our results support the viability of RL for QAS. For the Iris classification task, the RL-QAS agent exhibited stable learning behavior across all four evaluation metrics: accuracy, reward, gate count, and circuit depth. Compared to a random agent, the RL-QAS agent consistently discovered PQCAs with higher accuracy and lower complexity. The reward function—comprising equally weighted performance and complexity components—successfully guided the agent toward architectures with both high accuracy and low resource demands.

However, when applied to the more complex MNIST 2 dataset, the RL-QAS agent required significantly more effort to discover efficient PQCAs. While the agent demonstrated learning behavior and identified high-performing circuits, the training was unstable, and the agent failed to converge. These challenges underscore the need for enhanced search strategies to improve scalability and stability.

The \textit{Illegal Actions} mechanism proved effective in constraining the search space. By penalizing redundant or infeasible gate placements, the agent was encouraged to terminate episodes upon constructing an optimal PQCA, rather than continuing to the maximum allowed complexity. This behavior helped avoid overfitting and reduced unnecessary circuit depth. Importantly, the RL-QAS agent identified performant PQCAs while exploring only a small fraction of the total search space—demonstrating efficient learning and avoidance of unproductive architectural paths.

Across all Iris variants, the RL-QAS agent consistently discovered PQCAs achieving 100\% accuracy with minimal complexity. Compared to SEL VQCs, the RL-QAS circuits exhibited competitive or superior performance with substantially fewer gates and lower circuit depth—making them better suited for current NISQ devices. For the full Iris task and MNIST 2, RL-QAS VQCs outperformed single-layer SEL circuits in both performance and efficiency. Although a two-layer SEL circuit surpassed the RL-QAS circuit for MNIST 2 in accuracy, it did so at the cost of significantly increased complexity. Thus, RL-QAS VQCs offer a favorable trade-off between accuracy, efficiency, and practical implementability.

\section{Conclusion}
\label{sec:conclusion}

This work introduced RL-QAS, a RL-based framework for QAS that decouples architecture construction and parameter optimization into an outer and inner loop. Leveraging a tensor-based encoding scheme and a discrete gate set, the RL agent constructs candidate PQCAs while adhering to dynamic constraints enforced through an illegal action mechanism. A dual-objective reward function balances performance and complexity, favoring shallow, high-accuracy circuits. Efficient training is supported by a caching strategy that avoids redundant PQCA evaluations. Empirical validation on the Iris and MNIST 2 classification tasks demonstrated that RL-QAS consistently discovers compact, performant circuits that outperform SEL baselines in accuracy and architectural efficiency.

Despite these promising results, several challenges remain. The current evaluation is limited to two datasets and noise-free simulations; future work should explore more complex, unbalanced tasks and assess performance on actual quantum hardware. Enhancements such as integrating learned performance predictors, incorporating noise models, adapting to hardware constraints, and optimizing hyperparameters for the PPO algorithm could significantly improve scalability and robustness. Moreover, expanding the RL action space to include encoding strategies and alternative inner-loop optimizers presents an opportunity for more expressive and adaptable PQCAs. These directions will be crucial for extending RL-QAS to broader QML applications and real-world QC environments.

\appendix

\bibliography{main}

\clearpage



\begin{figure*}[htbp]
    \centering
    \subfloat[]{%
        \includegraphics[width=0.48\linewidth]{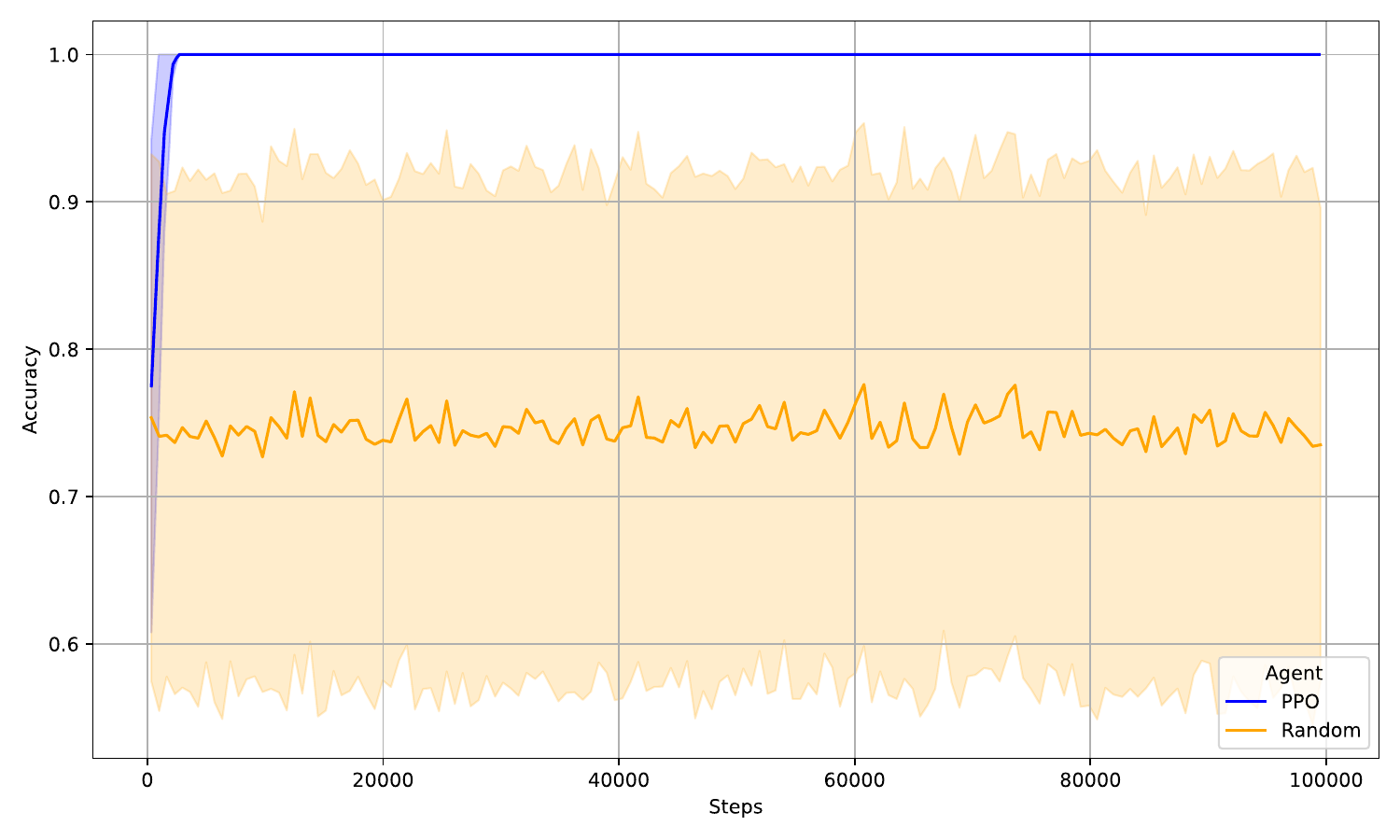}%
        \label{fig:accuracy_x_steps_iris_2_0-1}%
    }
    \hfill
    \subfloat[]{%
        \includegraphics[width=0.48\linewidth]{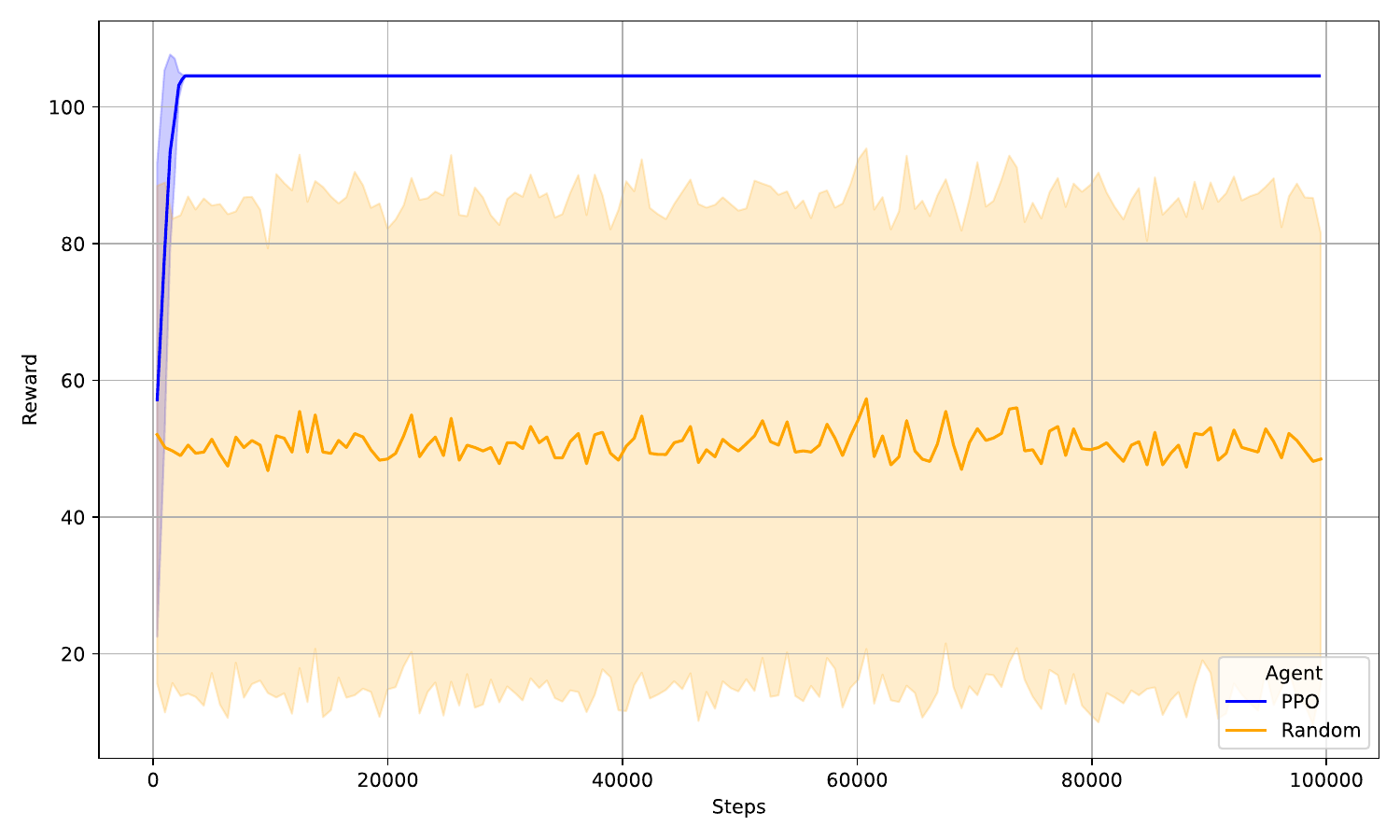}%
        \label{fig:reward_x_steps_iris_2_0-1}%
    }
    \caption{Achieved (a) accuracy and (b) reward as a function of the completed training steps for the Iris 2 (0, 1) classification problem}
    \label{fig:accuracy_reward_x_steps_iris_2_0-1}
\end{figure*}

\begin{figure*}[htbp]
    \centering
    \subfloat[]{%
        \includegraphics[width=0.48\linewidth]{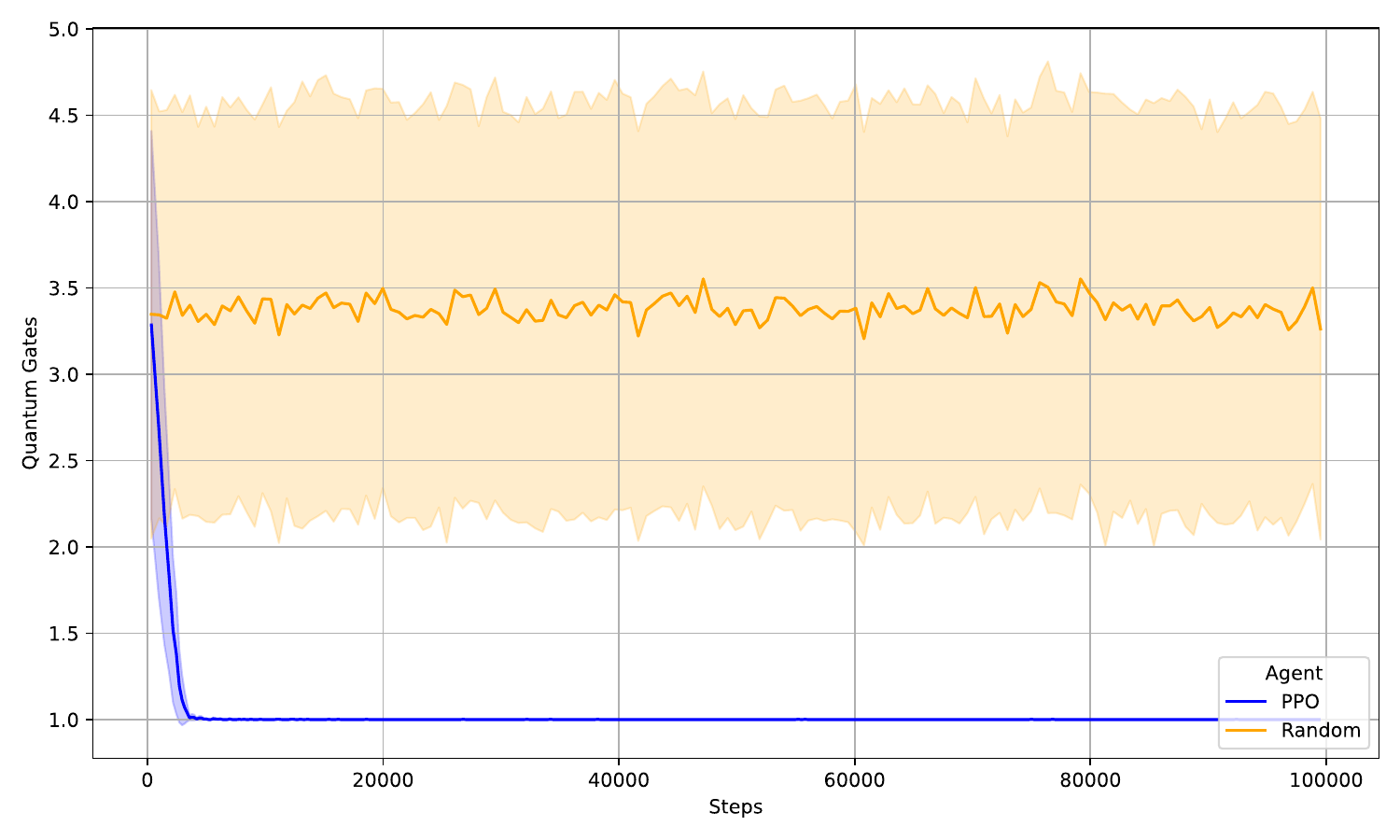}%
        \label{fig:gates_x_steps_iris_iris_2_0-1}%
    }
    \hfill
    \subfloat[]{%
        \includegraphics[width=0.48\linewidth]{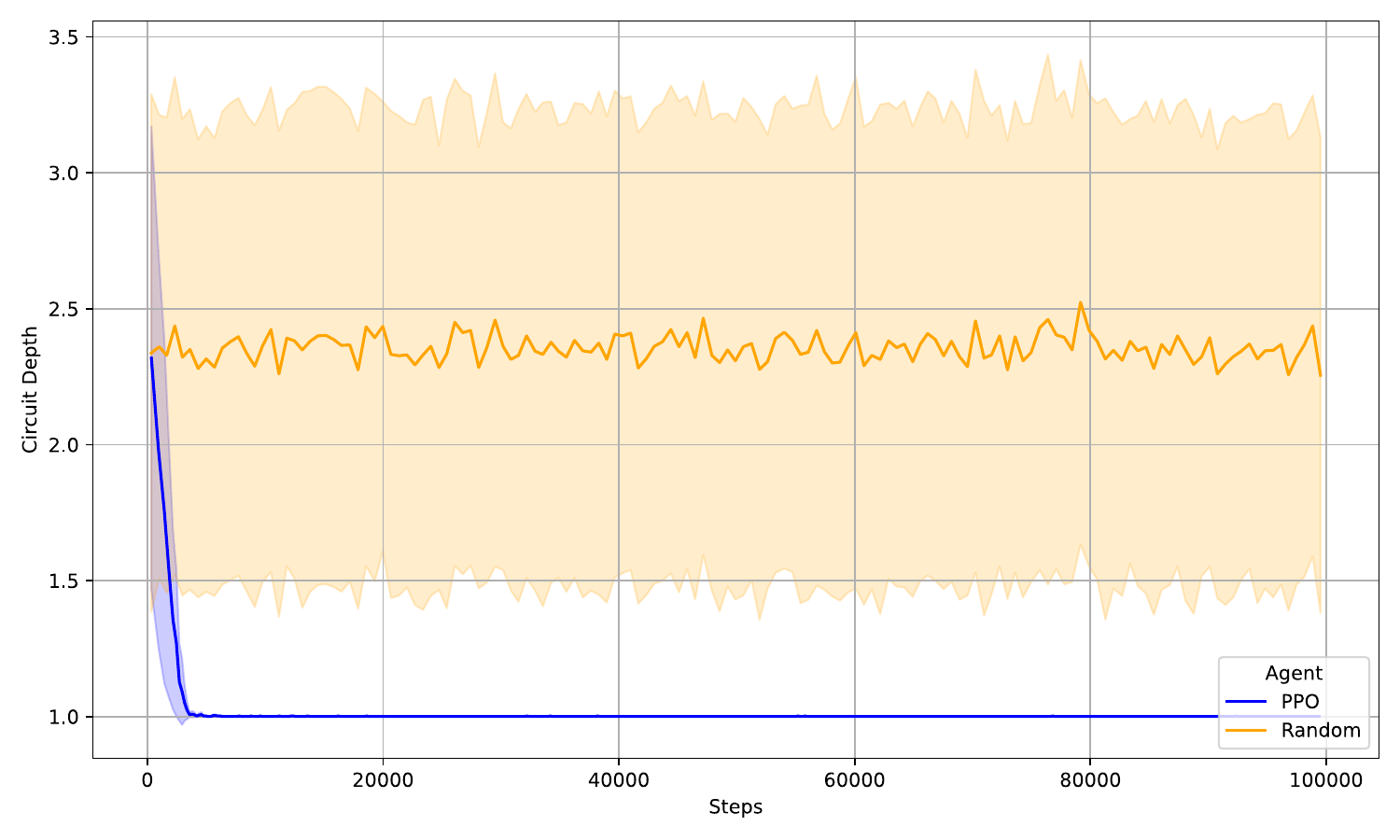}%
        \label{fig:depth_x_steps_iris_2_0-1}%
    }

    \caption{Number of (a) quantum gates used and (b) circuit depth utilized as a function of the completed training steps for the Iris 2 (0, 1) classification problem}
    \label{fig:gates_depth_x_steps_iris_2_0-1}
\end{figure*}

\begin{figure*}[htbp]
    \centering
    \subfloat[]{%
        \includegraphics[width=0.48\linewidth]{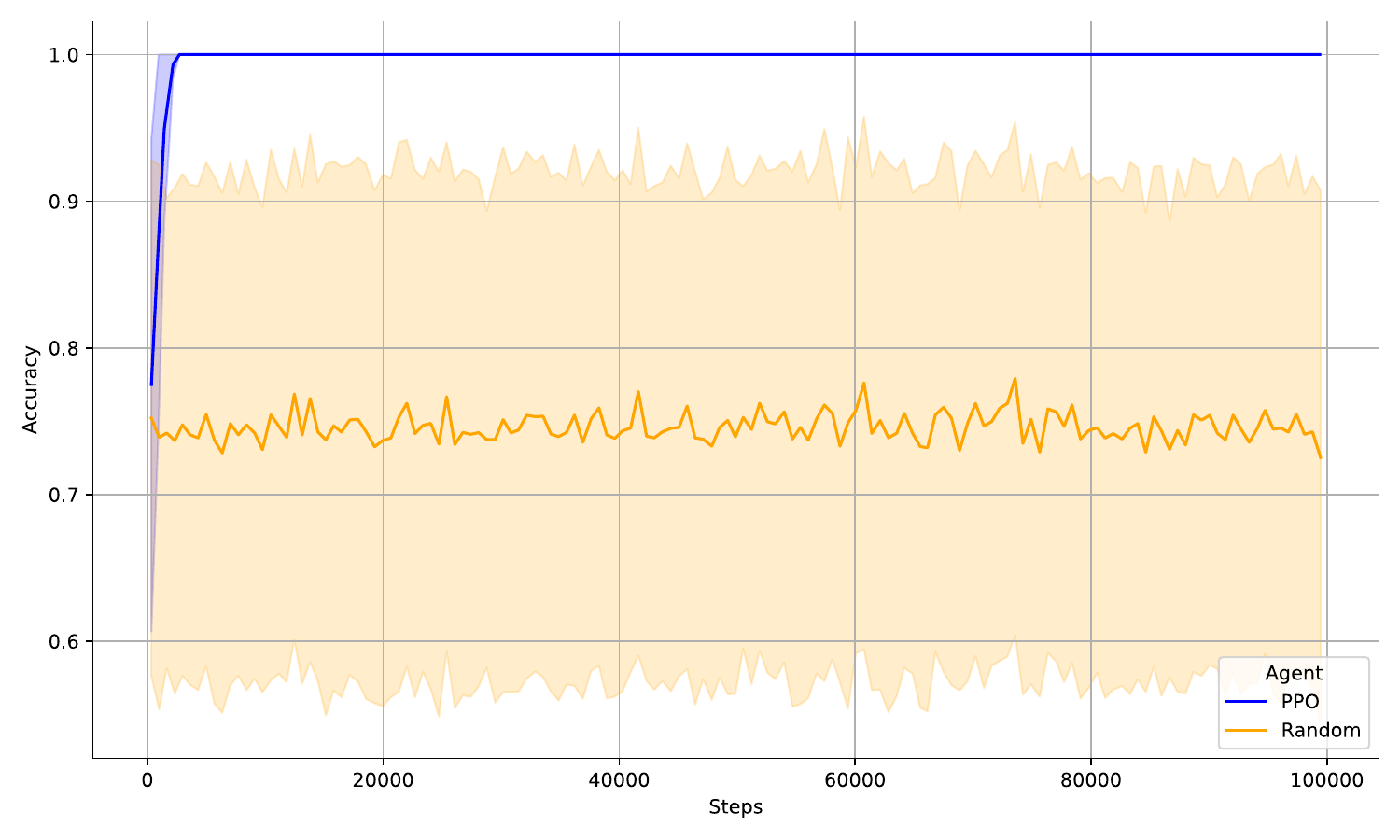}%
        \label{fig:accuracy_x_steps_iris_2_0-2}%
    }
    \hfill
    \subfloat[]{%
        \includegraphics[width=0.48\linewidth]{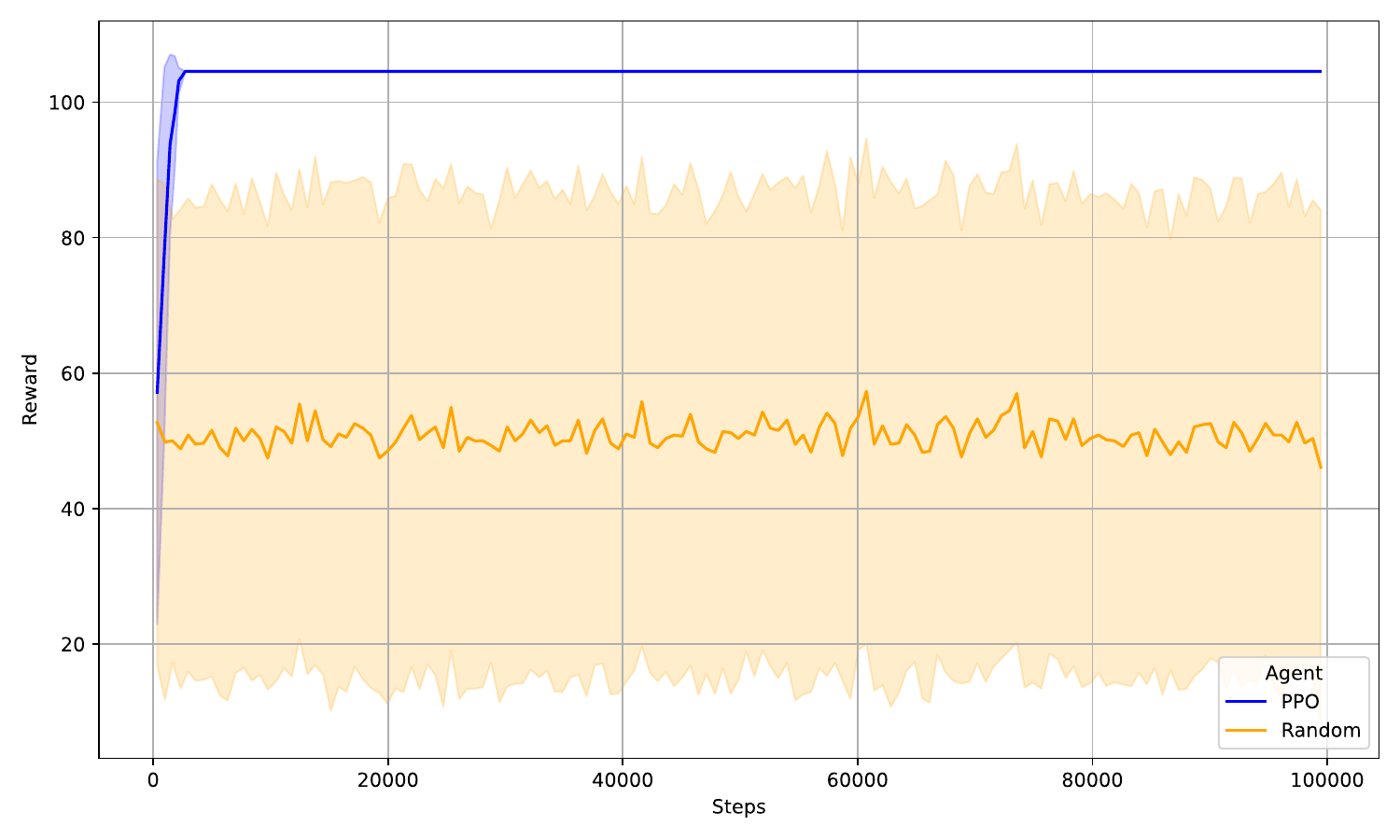}%
        \label{fig:reward_x_steps_iris_2_0-2}%
    }

    \caption{Achieved (a) accuracy and (b) reward as a function of the completed training steps for the Iris 2 (0, 2) classification problem}
    \label{fig:accuracy_reward_x_steps_iris_2_0-2}
\end{figure*}

\begin{figure*}[htbp]
    \centering
    \subfloat[]{%
        \includegraphics[width=0.48\linewidth]{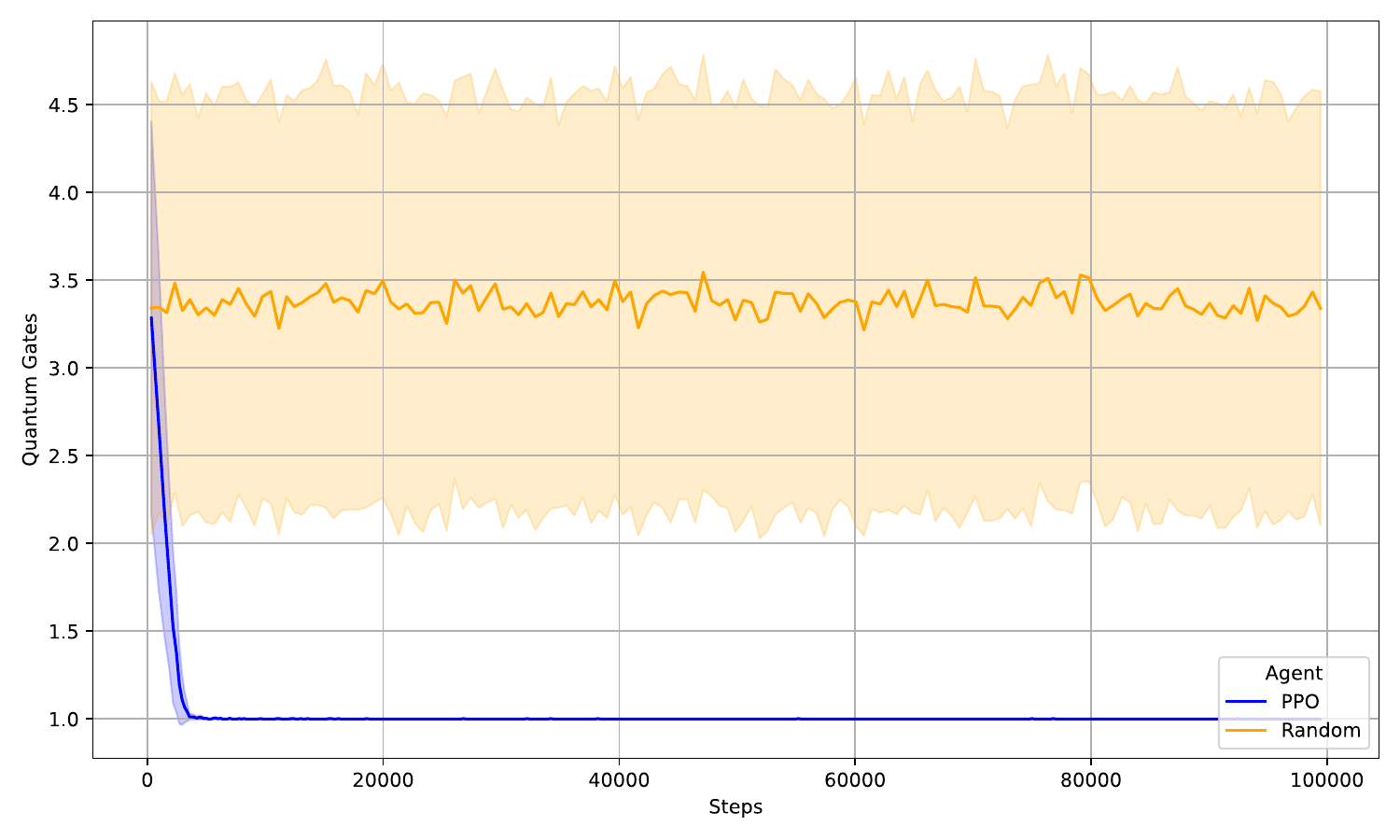}%
        \label{fig:gates_x_steps_iris_iris_2_0-2}%
    }\hfill
    \subfloat[]{%
        \includegraphics[width=0.48\linewidth]{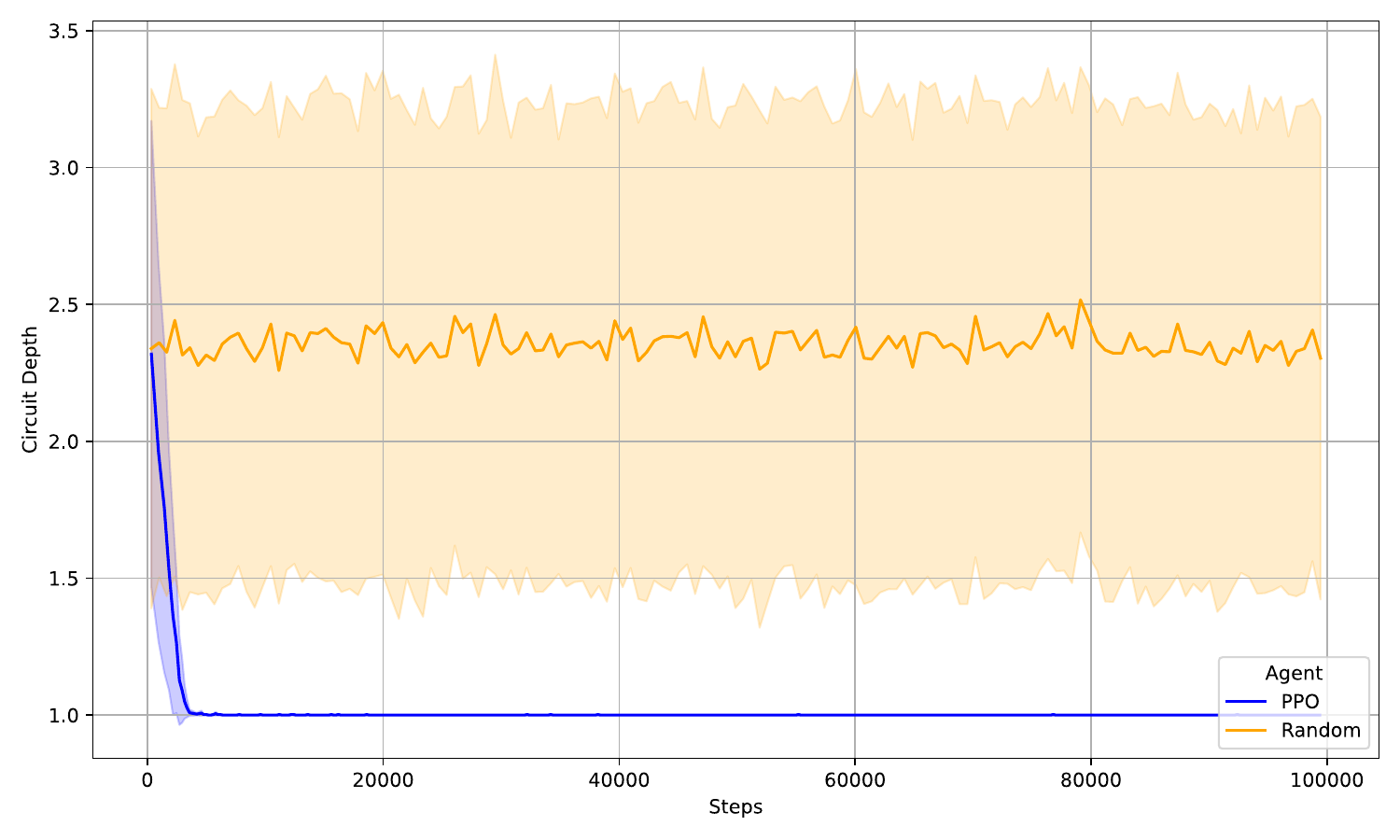}%
        \label{fig:depth_x_steps_iris_2_0-2}%
    }

    \caption{Number of (a) quantum gates used and (b) circuit depth utilized as a function of the completed training steps for the Iris 2 (0, 2) classification problem}
    \label{fig:gates_depth_x_steps_iris_2_0-2}
\end{figure*}

\begin{figure*}[htbp]
    \centering
    \subfloat[]{%
        \includegraphics[width=0.47\linewidth]{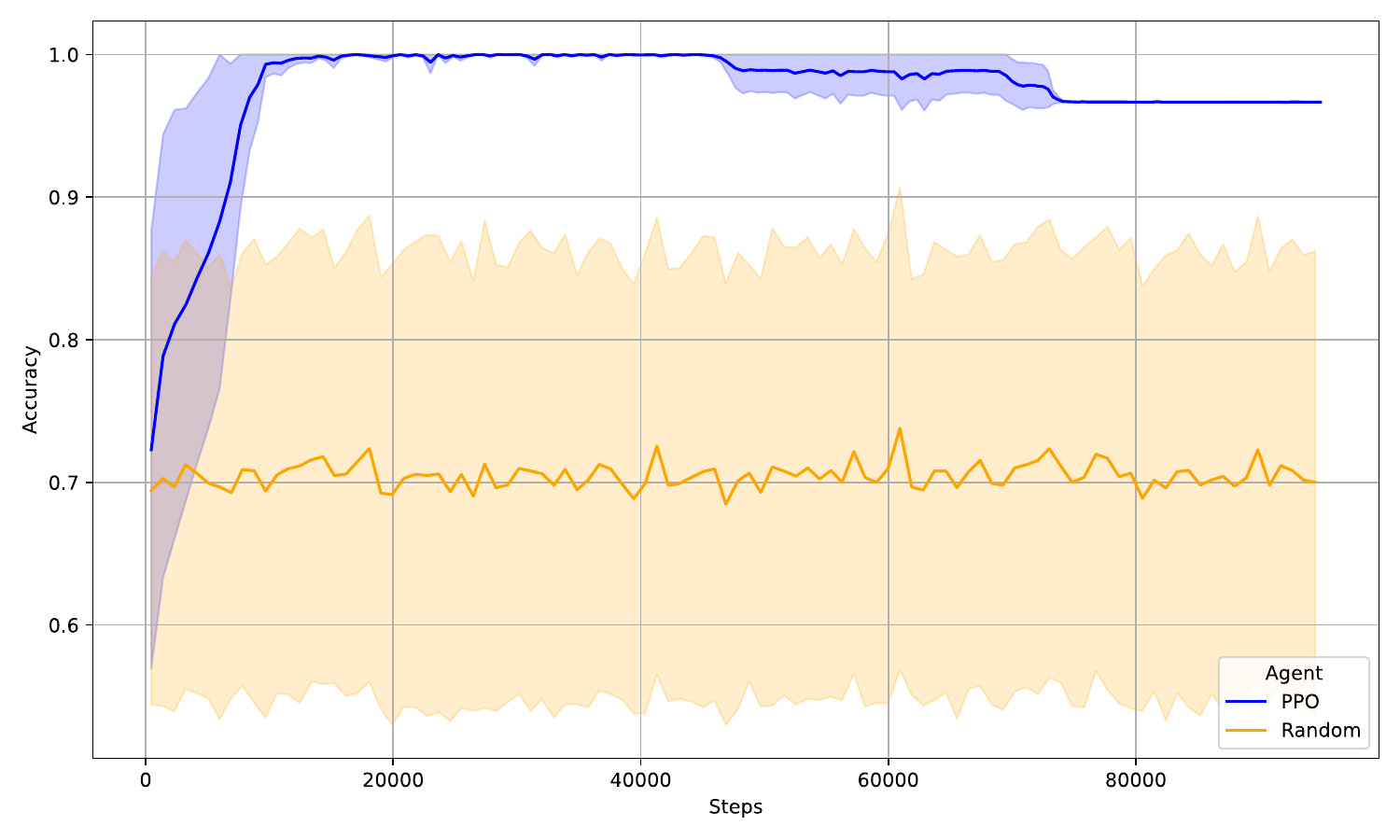}%
        \label{fig:accuracy_x_steps_iris_2_1-2}%
    }\hfill
    \subfloat[]{%
        \includegraphics[width=0.47\linewidth]{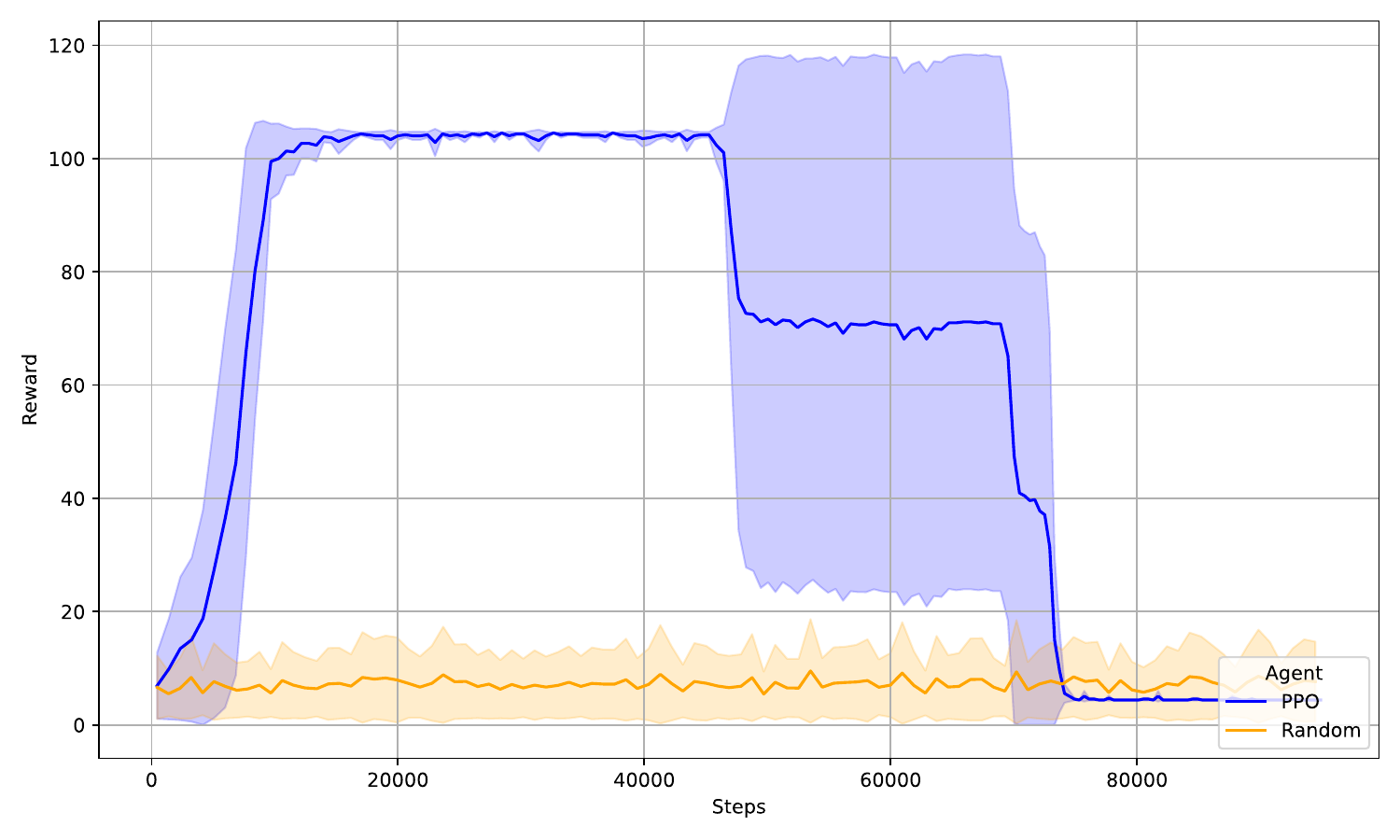}%
        \label{fig:reward_x_steps_iris_2_1-2}%
    }

    \caption{Achieved (a) accuracy and (b) reward as a function of the completed training steps for the Iris 2 (1, 2) classification problem}
    \label{fig:accuracy_reward_x_steps_iris_2_1-2}
\end{figure*}

\begin{figure*}[htbp]
    \centering
    \subfloat[]{%
        \includegraphics[width=0.48\linewidth]{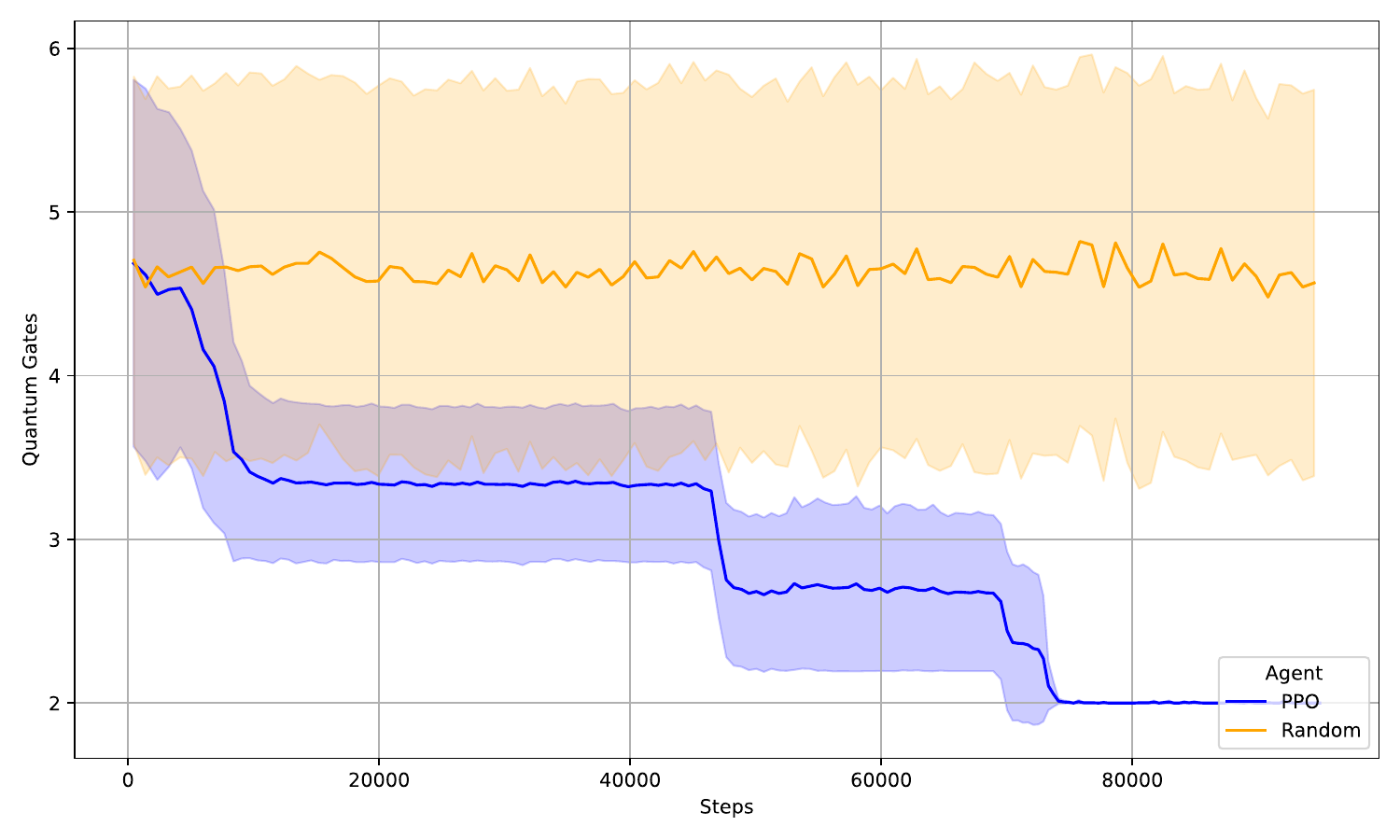}%
        \label{fig:gates_x_steps_iris_2_1-2}%
    }\hfill
    \subfloat[]{%
        \includegraphics[width=0.48\linewidth]{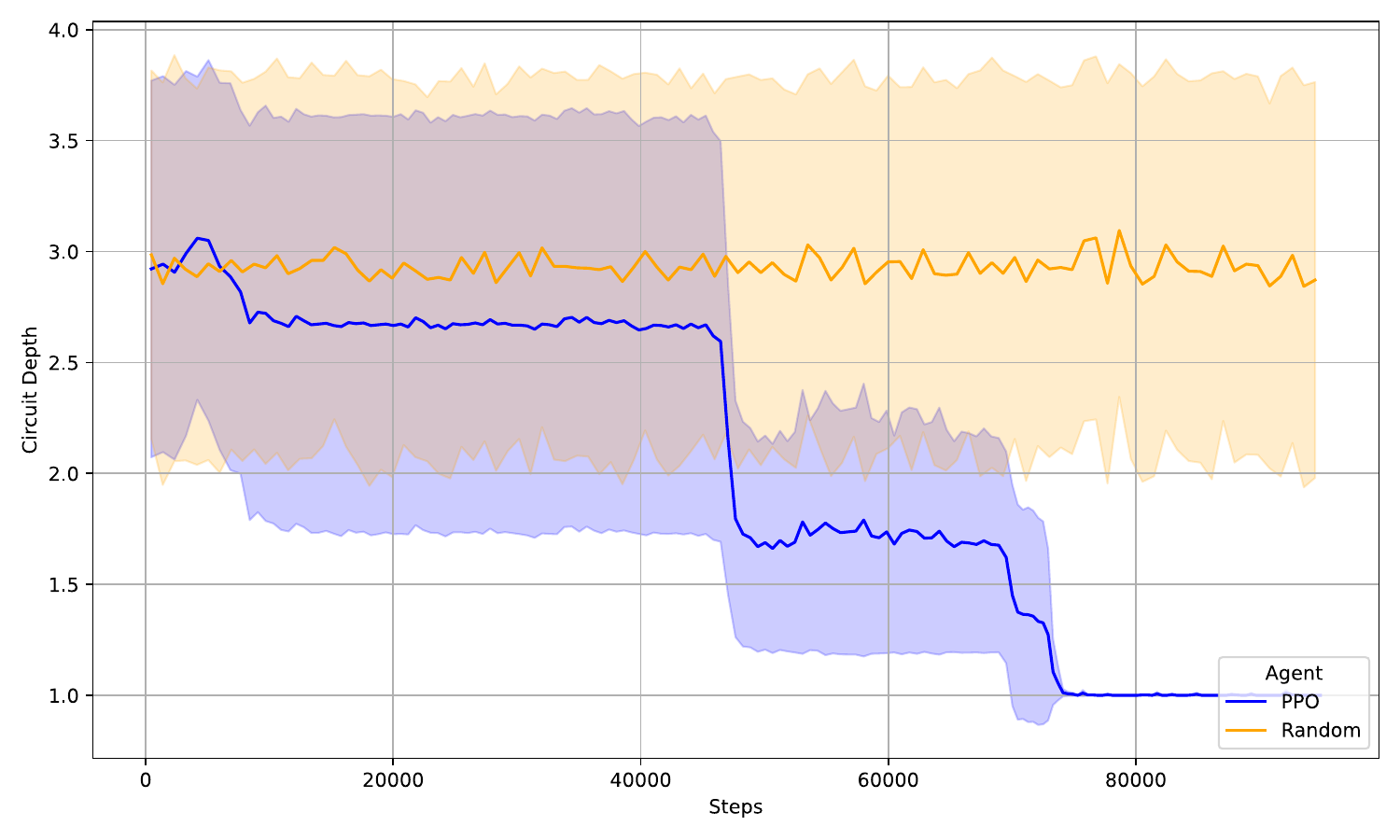}%
        \label{fig:depth_x_steps_iris_2_1-2}%
    }
    \caption{Number of (a) quantum gates used and (b) circuit depth utilized as a function of the completed training steps for the Iris 2 (1, 2) classification problem}
    \label{fig:gates_depth_x_steps_iris_2_1-2}
\end{figure*}

\begin{figure*}[htbp]
    \centering
    \subfloat[]{%
        \includegraphics[width=0.46\linewidth]{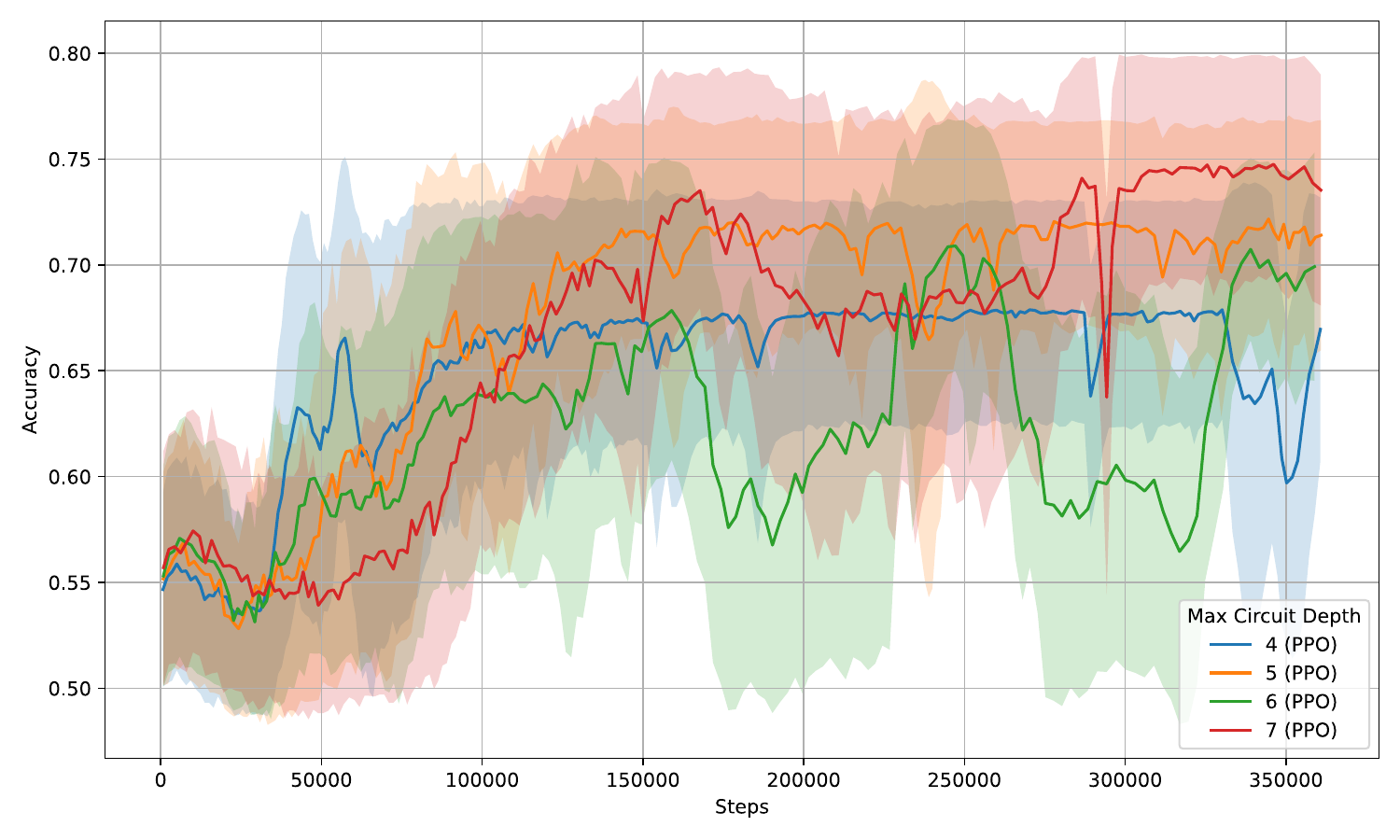}%
        \label{fig:accuracy_x_steps_mnist_2}%
    }
    \subfloat[]{%
        \includegraphics[width=0.46\linewidth]{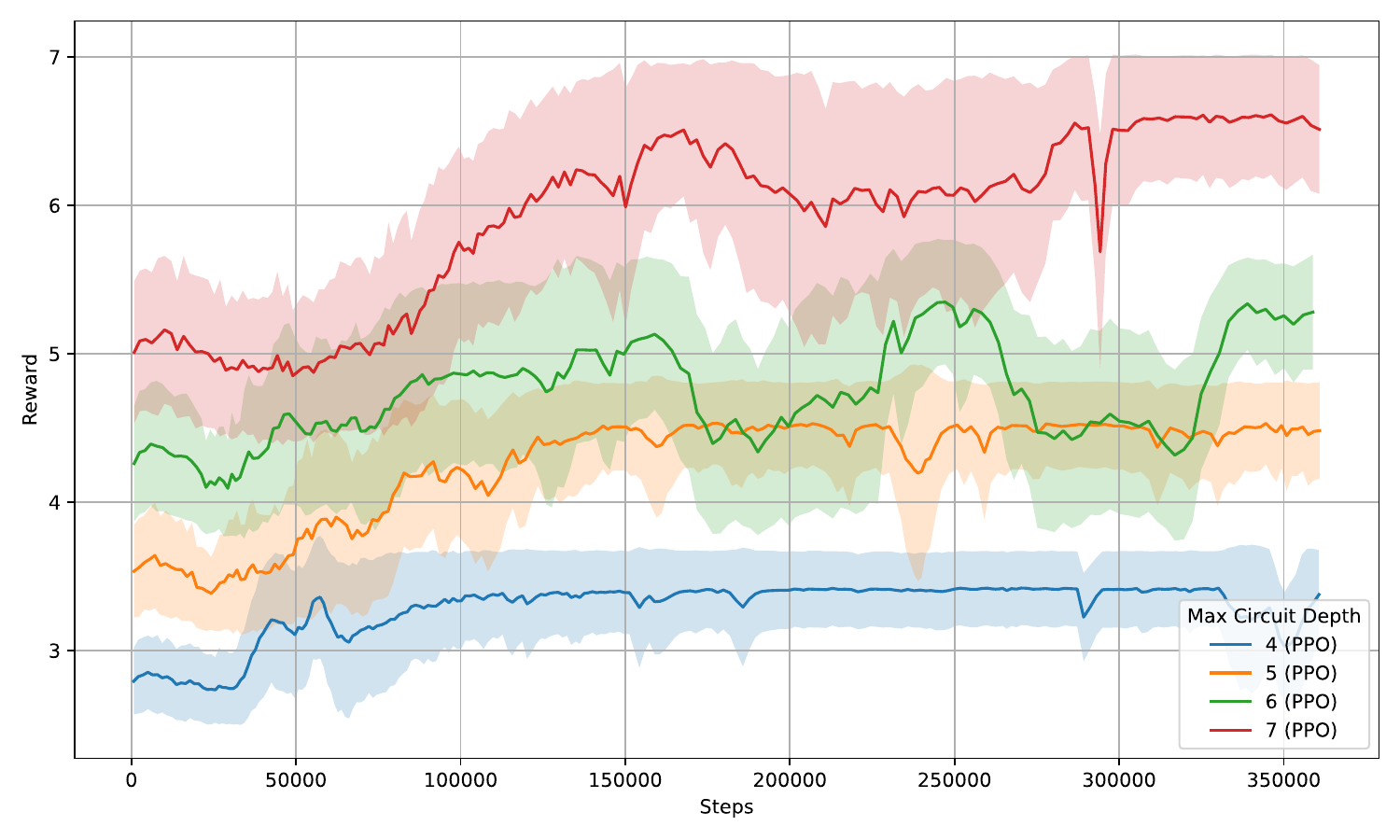}%
        \label{fig:reward_x_steps_mnist_2}%
    }

    \caption{Achieved (a) accuracy and (b) reward as a function of the completed training steps for the MNIST 2 classification problem}
    \label{fig:accuracy_reward_x_steps_mnist_2}
\end{figure*}

\begin{figure*}[htbp]
    \centering
    \subfloat[]{%
        \includegraphics[width=0.46\linewidth]{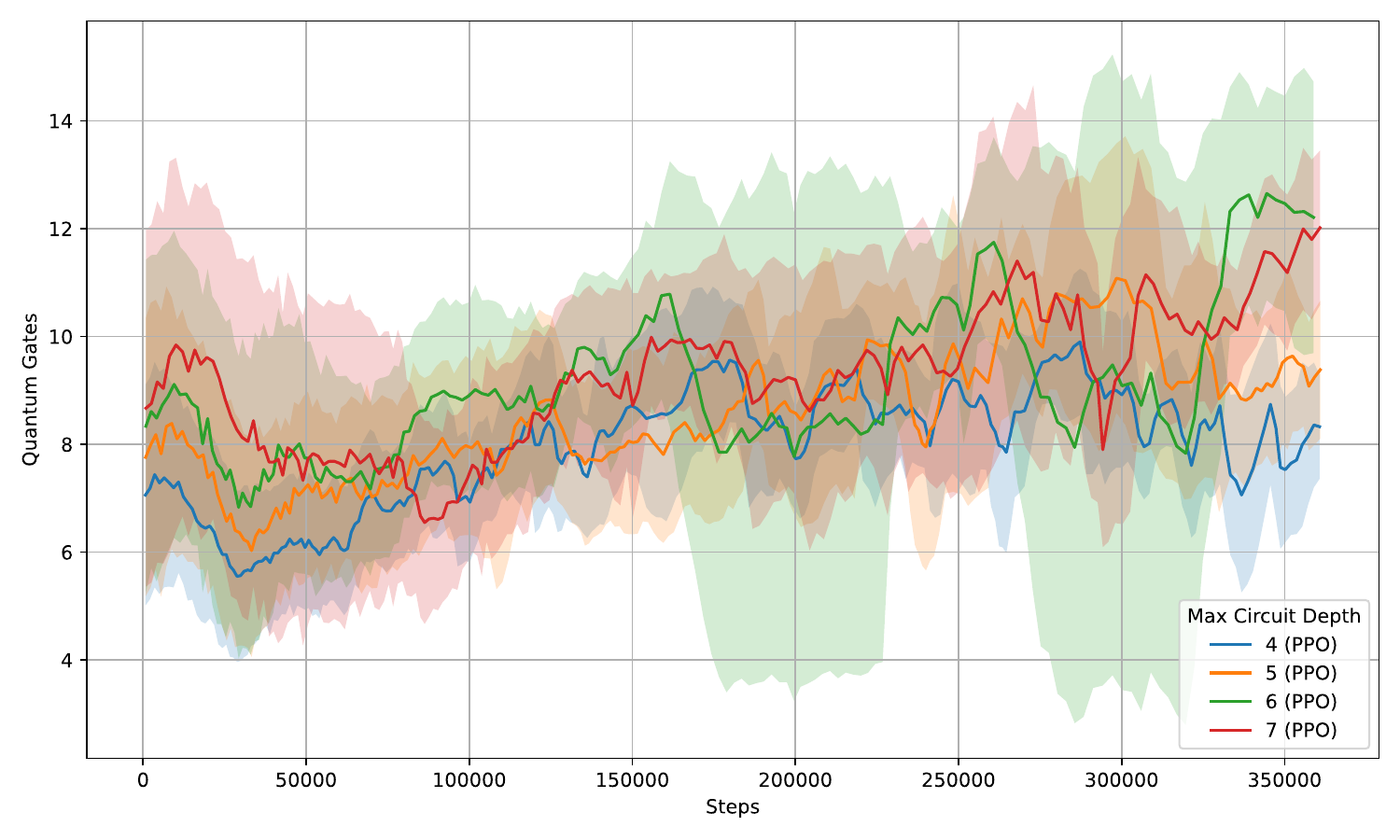}%
        \label{fig:gates_x_steps_mnist_2}%
    }
    \subfloat[]{%
        \includegraphics[width=0.46\linewidth]{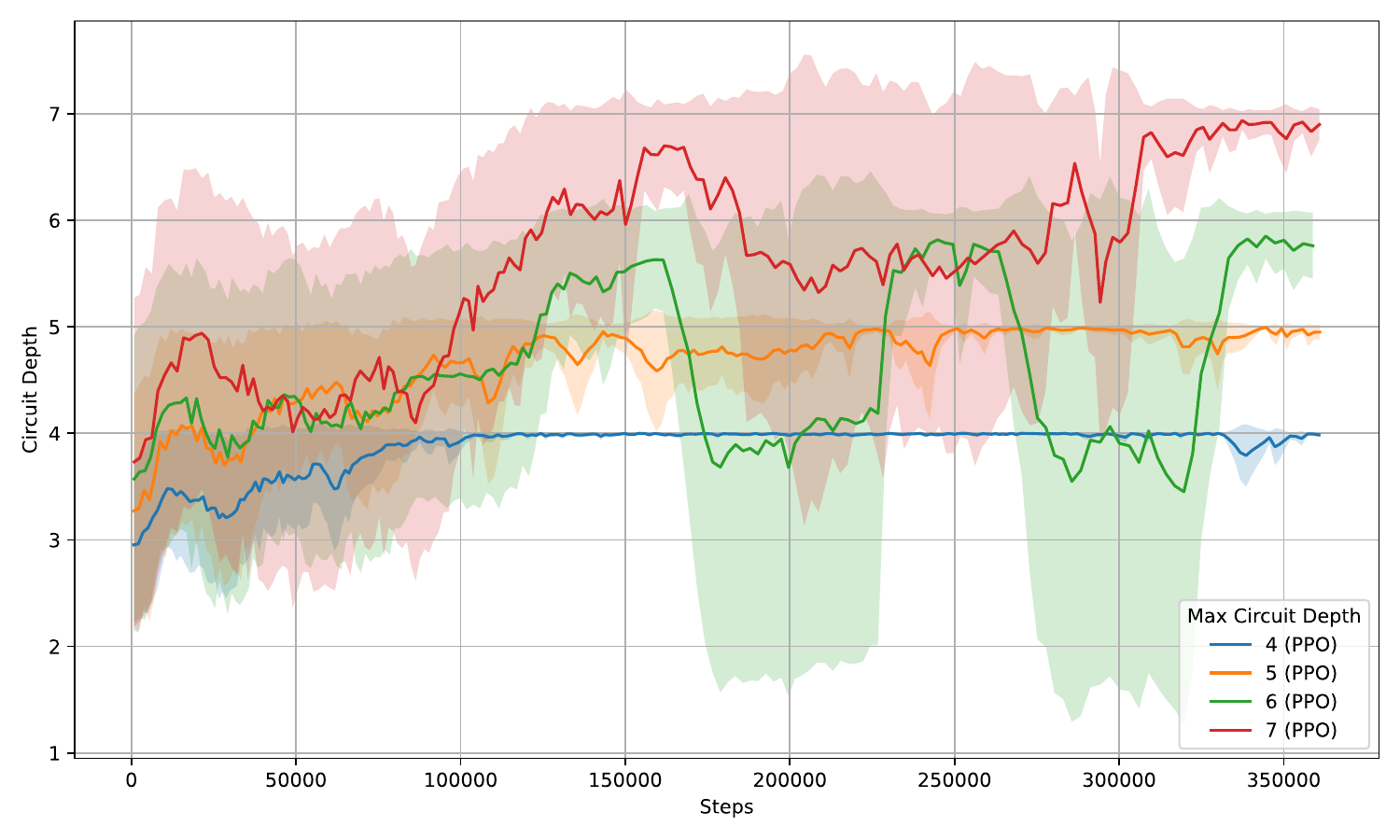}%
        \label{fig:depth_x_steps_mnist_2}%
    }

    \caption{Number of (a) quantum gates used and (b) circuit depth utilized as a function of the completed training steps for the MNIST 2 classification problem}
    \label{fig:gates_depth_x_steps_mnist_2}
\end{figure*}

\begin{figure*}[hpbt]
  \centering
  \subfloat[Iris 2 (0,1) classification problem]{
    \includegraphics[width=0.48\linewidth]{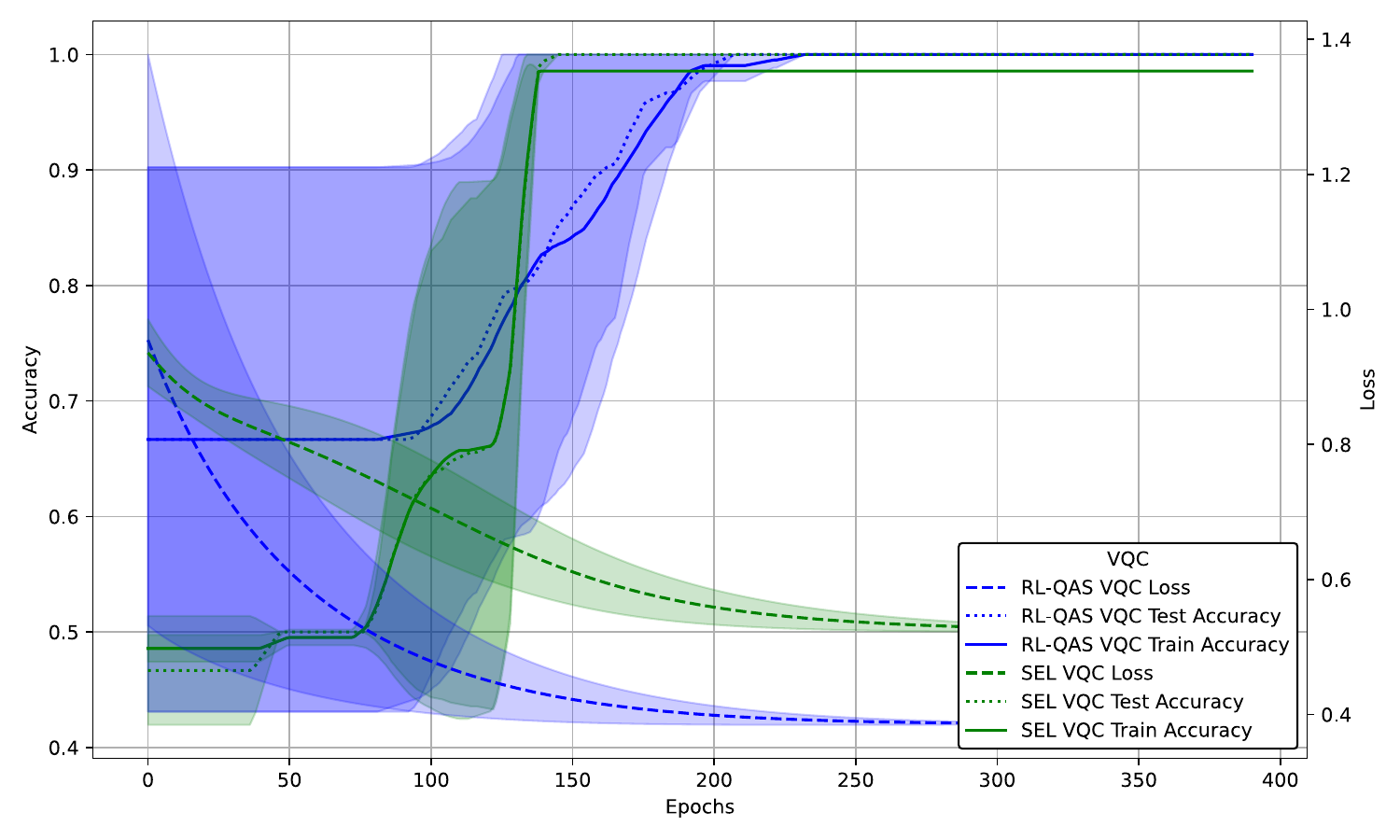}
    \label{fig:acc_loss_opt_epochs_iris_2_0-1_circuit1_1L_comparison}
  }
  \hfill
  \subfloat[Iris 2 (0,2) classification problem]{
    \includegraphics[width=0.48\linewidth]{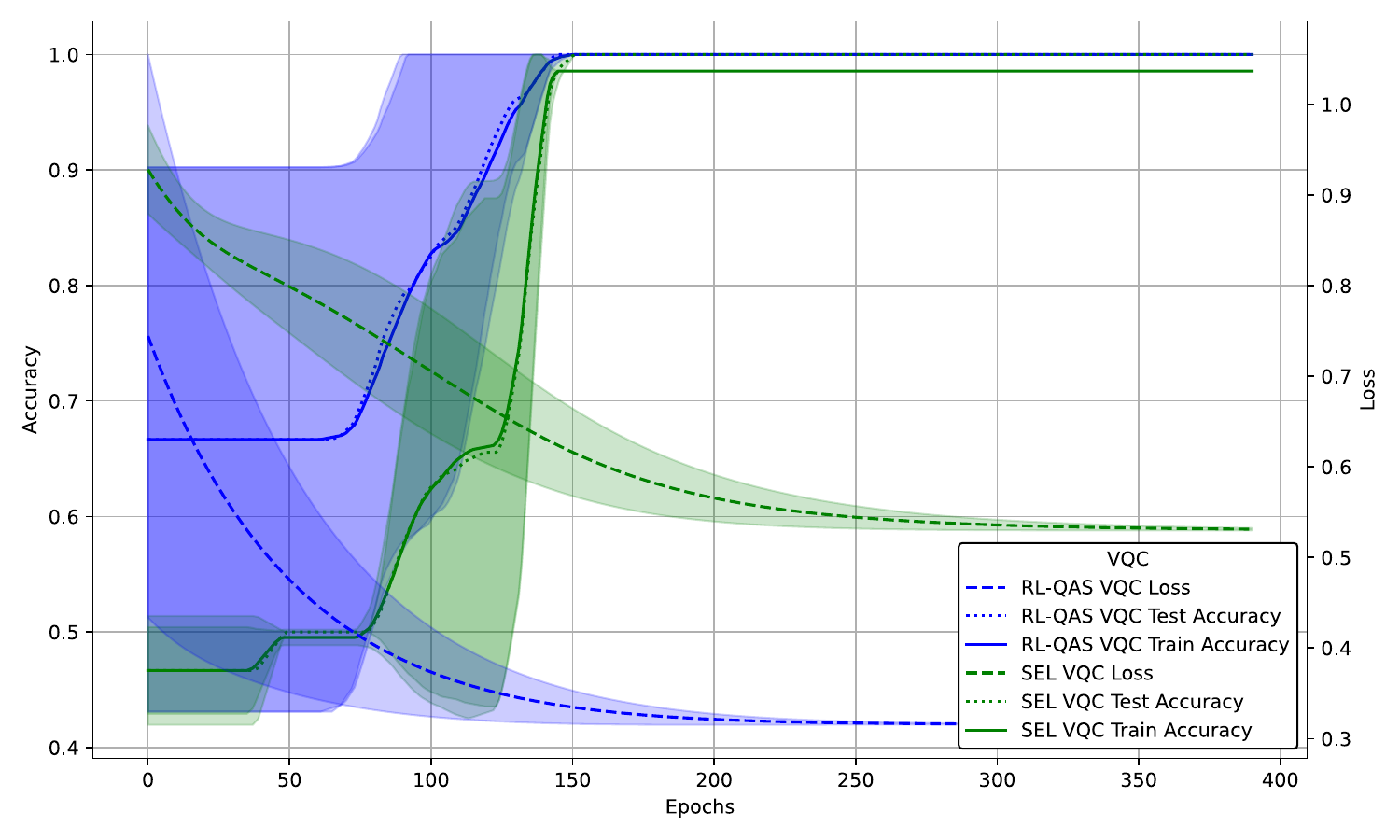}
    \label{fig:acc_loss_opt_epochs_iris_2_0-2_circuit1_1L_comparison}
  }

  \caption[Optimization behavior of the best PQCA for Iris 2 classification]{Optimization behavior of the best PQCA found within the RL-QAS for the Iris 2 classification problems \textbf{(a)}~(0,1) and \textbf{(b)}~(0,2), each compared to a SEL VQC with one layer.}
  \label{fig:acc_loss_opt_epochs_iris_2_comparison}
\end{figure*}

\begin{figure*}[hpbt]
  \centering
  \subfloat[Iris 2 (1,2) classification problem]{
    \includegraphics[width=0.48\linewidth]{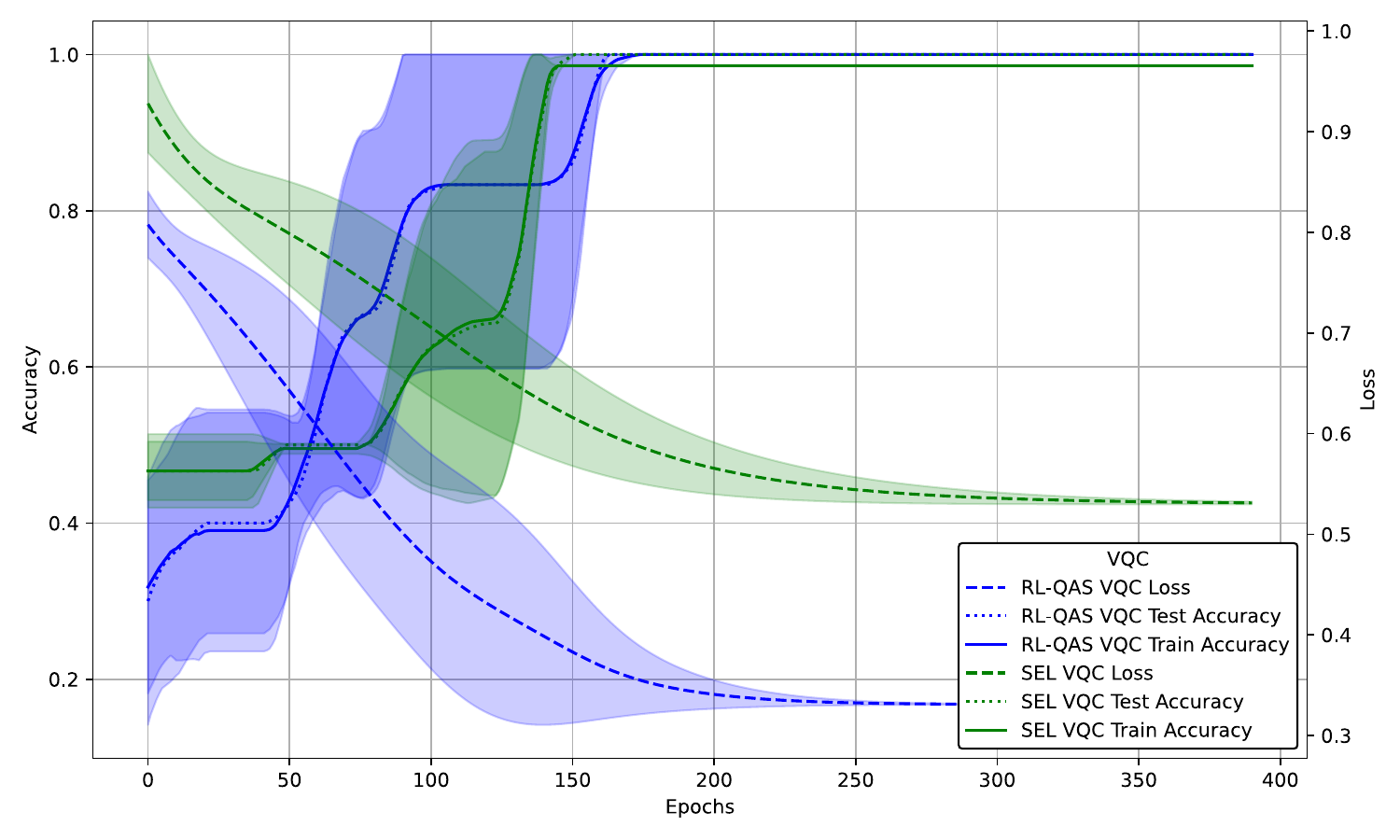}
    \label{fig:acc_loss_opt_epochs_iris_2_1-2_circuit1_1L_comparison}
  }
  \hfill
  \subfloat[MNIST 2 classification problem]{
    \includegraphics[width=0.48\linewidth]{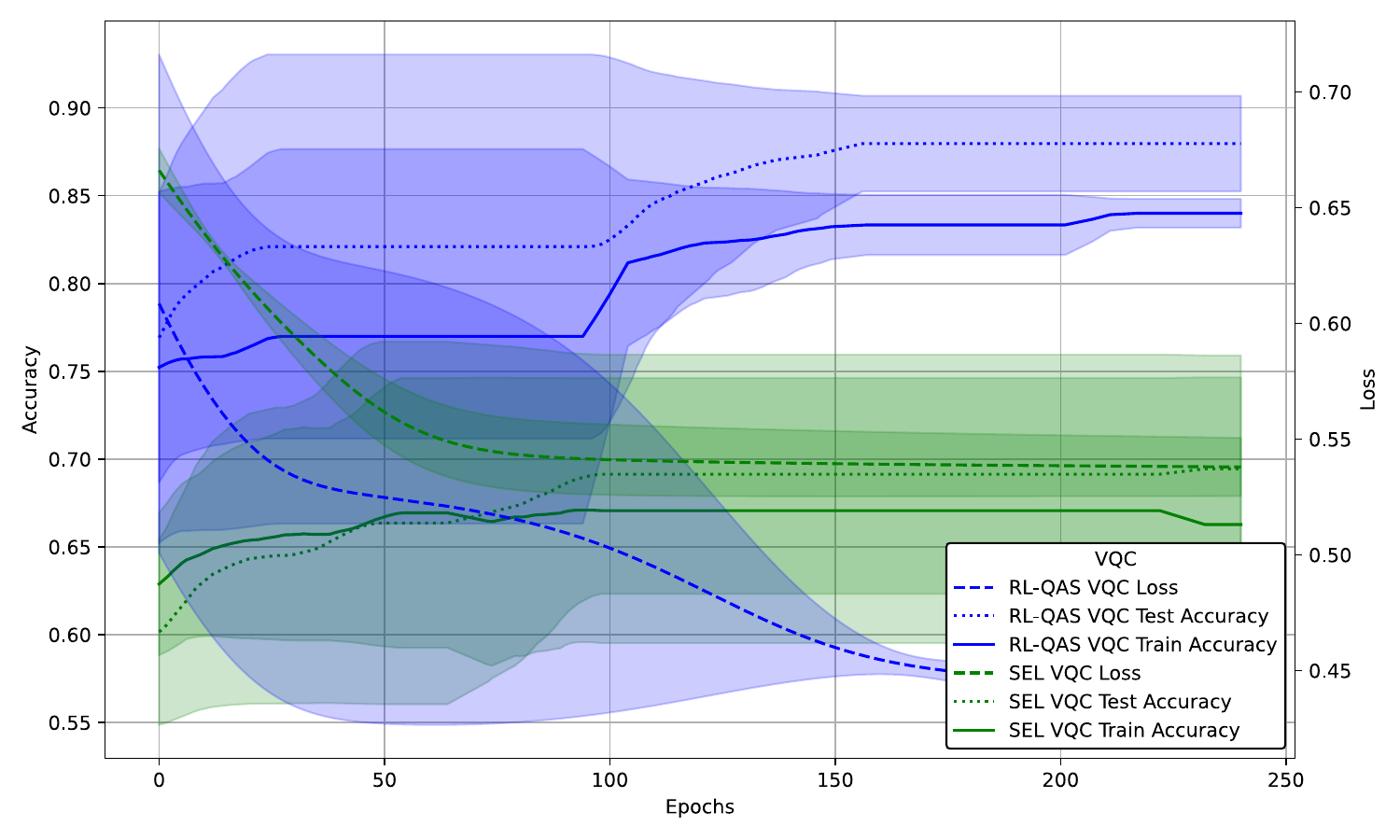}
    \label{fig:acc_loss_opt_epochs_mnist_2_circuit1_1L_comparison}
  }

  \caption[Optimization behavior of the best PQCA for Iris and MNIST classification]{Optimization behavior of the best PQCA found within the RL-QAS compared to a SEL VQC with one layer for the classification problems \textbf{(a)}~Iris 2 (1,2) and \textbf{(b)}~MNIST 2.}
  \label{fig:acc_loss_opt_epochs_iris_mnist_comparison}
\end{figure*}

\end{document}